\documentclass[a4paper,11pt]{article}

\usepackage{jcappub}          
\usepackage[utf8]{inputenc}
\usepackage{graphicx}
\usepackage{amsmath,amsfonts,amssymb,bm}
\usepackage{booktabs}
\usepackage{xcolor}
\usepackage{caption}
\usepackage{orcidlink}
\usepackage{microtype}

\newcommand{\PeV}{\,\mathrm{PeV}}
\newcommand{\TeV}{\,\mathrm{TeV}}
\newcommand{\GeV}{\,\mathrm{GeV}}
\newcommand{\kpc}{\,\mathrm{kpc}}
\renewcommand{\vec}[1]{\bm{#1}}

\title{Caught by its own light: \\Wimpzillas, the LHAASO knee \\and the diffuse $\gamma$-ray sky}

\author[a]{D.~Akl\orcidlink{0009-0006-4358-9929}}
\emailAdd{dalya.akl@nyu.edu}     
\author[a]{A.~J.~Iovino\orcidlink{0000-0002-8531-5962}}
\emailAdd{a.iovino@nyu.edu}
\author[a]{D.~Kantzas\orcidlink{0000-0002-7364-606X}}
\emailAdd{dimitrios.kantzas@nyu.edu}      
\author[a]{A.~V.~Macci\`o\orcidlink{0000-0002-8171-6507}}
\emailAdd{maccio@nyu.edu}    

\affiliation[a]{Center for Astrophysics and Space Science (CASS), New York University Abu Dhabi, PO Box 129188, Abu Dhabi, UAE}

\abstract{
We use the proton spectrum resolved by LHAASO across the knee to test superheavy dark matter in the nucleon channel at PeV energies. Both the hardening at $0.34\PeV$ and the knee at $3.3\PeV$ are reproduced by nucleons from a relic of mass $5\cdot 10^{7}\GeV$ decaying in the Galactic halo, with the mass as the only shape parameter of the spectrum and the background index profiled under an independent prior on the proton slope below $0.1\PeV$. It works equally well in all three hadronic-interaction-model reconstructions. On the other hand, the photons of the same cascade rule out this interpretation by about two orders of magnitude against the photon fraction of the cosmic radiation, and by about one order of magnitude against the LHAASO diffuse Galactic emission.
}

\begin{document}
\maketitle
\flushbottom

\section{Introduction}\label{sec:intro}
Using the hybrid operation of its wide-field Cherenkov telescope array together with the electromagnetic-particle and muon detectors of its square-kilometre array (KM2A), LHAASO has isolated a high-purity sample of primary protons and measured their spectrum from $0.158$ to $12.6\PeV$ \cite{LHAASO:2025byy}, and has done the same for helium above $0.1\PeV$ \cite{LHAASO:2025mlf}. The proton spectrum is not a simple broken power law. It hardens with respect to the extrapolation of the low-energy behaviour, with the spectral index moving from $\gamma_1 = 2.71 \pm 0.02 \pm 0.08$ to $\gamma_2 = 2.51 \pm 0.03 \pm 0.06$ across a break at $E_h = 0.34 \pm 0.02 \pm 0.04 \PeV$, and it then steepens to $\gamma_3 = 3.5 \pm 0.2 \pm 0.2$ above a knee at $E_k = 3.3 \pm 0.4 \pm 0.5 \PeV$. A hardening in the proton spectrum below the knee had already been reported by GRAPES-3, at around $166\TeV$ \cite{GRAPES-3:2024mhy}. Those numbers are the ones obtained with the EPOS-LHC reconstruction of the showers; the same fit is published for two further hadronic interaction models, giving $\gamma_1 = 2.79 \pm 0.07 \pm 0.12$ with QGSJETII-04 and $2.76 \pm 0.02 \pm 0.07$ with SIBYLL 2.3d, so even the slope of the component below the break is known only to within the choice of shower model.

The collaboration interprets this structure as the emergence of a cosmic-ray component that is distinct from the one dominating the flux below $0.1\PeV$, and tentatively associates it with the population of Galactic PeVatrons that LHAASO itself has been mapping. Minimal astrophysical descriptions of the proton and helium spectra from GeV to PeV energies have already been built on that basis \cite{Aharonian:2026tzf}, as have global fits connecting the Galactic and extragalactic components \cite{Lv:2024wrs}. The measurement has also been folded through hadronic production and tested against the diffuse Galactic $\gamma$-ray emission, which it overshoots \cite{Castro:2025wgf}. The hard component has recently attracted attention in the literature. It has been attributed to the sources, whether microquasars and the remnants of black-hole X-ray binaries \cite{Cooper:2020tzq, Kantzas:2023oww,Kaci:2025gyb,Zhang:2025tew,Zhang:2026igt}, the collective winds of young massive star clusters \cite{Qiu:2026kdu}, a spread in the maximum rigidity of the supernova remnant population \cite{Evoli:2026rpj} or the Cygnus region alone \cite{EspinosaCastro:2026xlr,Shi:2026unk}; and to transport, through a transition from diffusive to drift-dominated propagation \cite{EspinosaCastro:2026xbs} or a superposition of Galactic populations handing over early to the extragalactic component \cite{Yuan:2025xqt,Dzhatdoev:2026fvw}.

None of these explanations invokes physics beyond the Standard Model, though the array's photon data have been used for such a purpose several times. LHAASO has searched its whole sky for the bursts of evaporating primordial black holes and set the tightest limit on their local burst rate \cite{LHAASO:2025kyn}, a sensitivity to Hawking radiation forecast beforehand \cite{Yang:2024vij} and reconsidered in the light of the memory burden effect \cite{Dvali:2020wft,Alexandre:2024nuo,Chianese:2025wrk,Tan:2025vxp,Dondarini:2025ktz}. Its observation of GRB~221009A, both the $\TeV$ afterglow \cite{LHAASO:2023kyg} and the photons beyond $10\TeV$ \cite{LHAASO:2023lkv}, constrains axion-like particles \cite{Gao:2023und,Satunin:2025hbk} and sets some of the strongest available tests of Lorentz invariance violation \cite{LHAASO:2024lub}, complementing earlier tests using ultra-high-energy Galactic sources \cite{Cao:2021tat,Li:2021duv,Li:2024ivs}. Further, the diffuse emission has been used to constrain heavy and gravitationally produced dark matter \cite{LHAASO:2022yxw,Boehm:2025qro,Dubey:2025ouh,Barman:2025gjr, DeLaTorreLuque:2022PeV, DeLaTorreLuque:2025CRSea}.

Regardless of the origin of the feature, the measurement of LHAASO is a precise determination of the Galactic nucleon flux at PeV energies, and superheavy dark matter is predicted to contribute to exactly that flux. In the top-down scenario, a relic $\chi$ of mass far above the electroweak scale decays in the Galactic halo, and the resulting parton cascade delivers nucleons, photons, and neutrinos with a common, calculable spectrum \cite{Berezinsky:1997hy,Kuzmin:1997jua,Kuzmin:1998kk}. A relic this heavy cannot be thermal, since its mass lies far above the unitarity bound for a thermally produced species. It must instead be generated nonthermally, for example, through gravitational production during and after inflation, which can yield the observed abundance over a wide range of masses \cite{Chung:1998zb,Chung:1998ua,Chung:2001cb,Kolb:2023ydq}, as in the wimpzilla scenario~\cite{Kolb:1998ki}. 
A single decay cascade produces the photons and the nucleons together, in a ratio fixed by the fragmentation functions \cite{Berezinsky:1997hy}, so the nucleon flux is a prediction of such decay. That channel has not been tested at PeV energies, and dark matter decay has not been considered as a source of the knee\footnote{The one systematic study that uses measured cosmic-ray hadrons to bound decaying heavy dark matter \cite{Ishiwata:2019aet} employs Auger protons above $10^{9}\GeV$ and AMS-02 antiprotons below $10^{2}\GeV$, leaving the seven decades in between untouched. Every other constraint in the $10^{6}$ to $10^{15}\GeV$ mass range comes from $\gamma$-rays or neutrinos \cite{Kalashev:2016cre,Cohen:2016uyg,Chianese:2021jke,LHAASO:2022yxw,Rocamora:2025ddt,Boehm:2025qro,Dubey:2025ouh,Kachelriess:2018rty,Aloisio:2025nts}, or, for the very heaviest relics, from the ultra-high-energy photon and composition data of the Pierre Auger Observatory \cite{Alcantara:2019sco,Das:2023wtk,PierreAuger:2025jwt}. Dark matter has entered the knee literature through only one mechanism, the scattering of cosmic rays on halo particles, which needs a cross section of order a millibarn together with a large dark matter density in the disc \cite{Masip:2008mk,Masip:2009bk,Barcelo:2009uy,Tomozawa:2010we}.}.

We close that gap here. Section~\ref{sec:setup} sets up the two observables, with the fragmentation spectra of Ref.~\cite{Bauer:2020jay} and the diffusion equation solved in a slab halo. Section~\ref{sec:fit} confronts the proton measurement with the decay spectrum. Section~\ref{sec:gamma} follows the same decay into the photon channel, where the LHAASO diffuse Galactic emission \cite{LHAASO:2023gne} and the photon fraction of the cosmic radiation exclude the interpretation independently. Section~\ref{sec:bound} turns both into lifetime bounds under a single criterion. We conclude in Section~\ref{sec:concl}.
\section{Nucleons and photons from a decaying relic}\label{sec:setup}
\subsection{The decay spectrum}
When a particle of mass $m_\chi \gg m_Z$ decays into coloured partons, the resulting cascade is governed by the same physics that controls jet fragmentation at colliders, evolved over a very much larger range of scales, with electroweak radiation becoming as important as QCD radiation once $\alpha_2 \ln^2(m_\chi/m_W)$ exceeds unity \cite{Ciafaloni:2010ti}. We take the spectra from \texttt{HDMSpectra} \cite{Bauer:2020jay}, which evolves the full Standard Model fragmentation functions from the mass of the decaying particle down to the electroweak scale and matches to a conventional shower below it, and which tabulates the yields of photons, neutrinos, electrons and positrons, and protons and antiprotons, for parent masses up to the Planck scale. We write $x = 2E/m_\chi$ throughout, so that $x = 1$ is the kinematic endpoint of a two-body decay $\chi \to X\bar X$.
\subsubsection{Energy budget} The rest mass of the relic is converted into a parton cascade, and the question is how much of it ends up in each stable species. Integrating the tabulated yields for $\chi \to b\bar b$ at $m_\chi = 5\cdot 10^{7}\GeV$, nucleons and antinucleons together carry a fraction $f_N = 0.088$ of the parent mass and photons a fraction $f_\gamma = 0.260$, so the cascade delivers about three times as much energy in photons as in nucleons. The reason is hadronisation independently of the dark matter model. A fragmenting quark jet produces mostly pions, of which isospin makes roughly a third neutral, and each neutral pion converts promptly into two photons; nucleons instead require baryon production, which the fragmentation process suppresses because building a baryon costs a diquark pair from the vacuum where a meson costs only a quark pair. The photon channel is therefore favoured by the strong interaction itself, and the excess is an energy excess and not merely a counting one. Indeed, photons outnumber nucleons by about eight to one in the range we integrate, but that number keeps growing as the integration reaches further down in $x$, because photons are the softer species, whereas the energy ratio is stable at $2.9$ whether the integral starts at $x = 10^{-6}$ or at $10^{-4}$.
That ratio is also insensitive to everything the model leaves free. Fragmentation functions evolve only logarithmically with the hard scale, so $f_\gamma/f_N$ stays between $2.8$ and $3.3$ over twelve decades of parent mass, and it stays between $2.3$ and $3.6$ across the six decay channels of Sec.~\ref{sec:fit}, where we fix the extremes of integrations to be $gg$ and $W^+W^-$.

\begin{figure}[t]\centering \includegraphics[width=\textwidth]{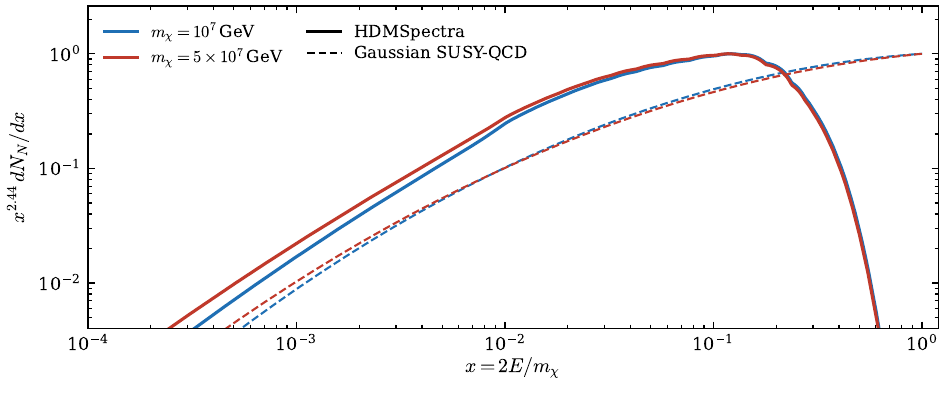}
\caption{The nucleon yield per decay, weighted by $x^{2.44}$, which is the combination that controls the size of a decay contribution relative to a cosmic-ray background of index $2.77$ propagated with $\tau_{\rm eff}\propto E^{-0.33}$, the index fitted in Sec.~\ref{sec:fit} and the diffusion index of the baseline setup above the rigidity break. Solid curves are \texttt{HDMSpectra} \cite{Bauer:2020jay}; dashed curves are the Gaussian SUSY-QCD spectrum of Refs.~\cite{Berezinsky:1997hy,Berezinsky:1998ed}. Each curve is normalised to its own peak.}\label{fig:frag}\end{figure}

\subsubsection{Shape at large $x$} The analytic form used in essentially all top-down work, the Gaussian limit of the SUSY-QCD cascade \cite{Berezinsky:1997hy,Berezinsky:1998ed},
\begin{equation}\label{eq:gauss} W_h(x) \;\propto\; \frac{1}{x}\,\exp\!\left[-\,\frac{\ln^2(x/x_m)}{2\sigma^2}\right], \qquad 2\sigma^2 = \frac{1}{6}\left[\ln\frac{m_\chi}{\Lambda}\right]^{3/2}, \qquad x_m = \left(\frac{\Lambda}{m_\chi}\right)^{1/2}, \end{equation}
with $\Lambda = 0.234\GeV$, is derived for $x \ll 1$ and is extrapolated to $x \sim 1$ and truncated by hand\footnote{In Ref.~\cite{Berezinsky:1998ed} the same spectrum is written in terms of $x = 2E/\sqrt{s}$ and with $\Lambda \simeq 0.25\GeV$, and its width reduces to the one in Eq.~\eqref{eq:gauss} for $b_{\rm SUSY} = 3$. We use throughout the convention and the value of $\Lambda$ of Ref.~\cite{Berezinsky:1997hy}.}. Figure~\ref{fig:frag} compares the two. They agree at small $x$ but differ qualitatively above $x \sim 10^{-2}$, where the analytic form has no endpoint suppression and produces a plateau terminated by an artificial cut at $x = 1$, while the true fragmentation function falls steeply already at $x \simeq 0.3$. 

Moreover, since an air shower initiated by an antiproton is not distinguished from one initiated by a proton, the quantity that LHAASO calls the proton flux is really the sum of protons and antiprotons, and we compare the data with the full nucleon yield throughout.
\subsection{Propagation of halo-produced nucleons}\label{sec:prop}
Decays of the halo population produce nucleons at a rate per unit volume and per unit energy
\begin{equation}\label{eq:source} Q(\vec{r},E) \;=\; \frac{\rho_\chi(\vec{r})}{m_\chi \tau_\chi}\, \frac{dN_N}{dE} , \end{equation}
where $\tau_\chi$ is the lifetime and $\rho_\chi$ the dark matter density, which we take to be an NFW profile \cite{Navarro:1996gj} with the parameters adopted by LHAASO in its own dark matter search, namely $\rho_s = 0.33\GeV\,\mathrm{cm}^{-3}$, $r_s = 20\kpc$, and $R_\odot = 8.3\kpc$, corresponding to a local density $\rho_\odot = 0.4 \GeV\,\mathrm{cm}^{-3}$ \cite{LHAASO:2022yxw}. Protons of PeV energy are still magnetically confined, their Larmor radius being a few parsecs in a microgauss field. The flux that reaches the Earth is therefore set by how effectively they are confined in the halo, and not by an integral along the line of sight as in the case of photons. 

In a slab halo of half-height $H$ and radius $R$, with a spatially uniform diffusion coefficient $K(E)$, as in the standard frameworks \cite{Strong:2007nh,Evoli:2016xgn}, and free escape at the boundary, the steady-state density obeys 
\begin{equation}
-K\,\nabla^2 n = Q
\end{equation}with the source of Eq.~\eqref{eq:source}. Writing $-\nabla^2 \psi = \rho_\chi/\rho_\odot$ with $\psi = 0$ on the boundary we get
\begin{equation}\label{eq:Jp} J_N(E) \;=\; \frac{c}{4\pi}\, \tau_{\rm eff}(E)\, \frac{\rho_\odot}{m_\chi \tau_\chi}\, \frac{dN_N}{dE} , \qquad \textrm{with} \qquad \tau_{\rm eff}(E) = \frac{\psi_\odot}{K(E)} , \end{equation}
and $\psi_\odot = \psi(R_\odot, 0)$ being a purely geometrical quantity with the dimensions of area. Indeed, the effective time $\tau_{\rm eff} = \psi_\odot/K$ is the density maintained at the Sun per unit local source rate, and $\psi_\odot$ is what remains of it once the diffusion coefficient is divided out. It therefore has the dimensions of an area, and it depends only on where the sources are and where the boundaries lie.

We solve for $\psi$ on a cylindrical grid, as set out in Appendix~\ref{app:diffusion}, where the simplest case fixes the scale: for a source spread uniformly through the slab, the solution is a parabola in $z$, with $\psi_\odot = H^2/2$. $H^2$ behaves like a random walk, namely a particle leaves once it has diffused a distance $H$, so the time it spends inside, and with it the density it maintains, grows as the square of the half-height.

The halo source is not uniform, but it is close to uniform as long as the box is thin compared with the scale on which the profile varies. For the baseline NFW profile, whose scale radius is $20\kpc$, we obtain $\psi_\odot = 8.7\kpc^2$ at $H = 4\kpc$ and $34.7\kpc^2$ at $H = 10\kpc$. The first is within ten per cent of $H^2/2$, because over a slab that thin the NFW density barely varies and the source is effectively uniform. The second is a third below it, because a box of that height already reaches out to where the profile has begun to fall.

A disc source behaves quite differently. If the injection is confined to $|z| < h$ with $h = 0.1\kpc$, the same calculation returns $hH$ in place of $\psi_\odot$, which is $1.0\kpc^2$ at $H = 10\kpc$. The halo source is therefore $35$ times more effective, and the origin of that factor is the ratio of the two thicknesses: for a uniform source, it would be exactly $H/2h = 50$, and the fall-off of the halo profile inside the box reduces it to $35$. A halo source injects particles all along the escape path, whereas a disc source injects them only in a layer a hundred times thinner, so almost the entire volume contributes nothing.

We stress that $\psi_\odot$ and $hH$ are the same quantity computed for two source geometries, and not two different physical times. Importing a confinement time calibrated on disc-injected nuclei \cite{Loeb:2002ee} therefore underestimates a halo signal by that factor of $35$ from the geometry alone, before the energy dependence of the disc-calibrated coefficient is taken into account. The last row of Table~\ref{tab:prop} reports that case for comparison.

It is natural to ask at this point how much of what follows depends on the halo profile, since a cored profile changes the source distribution inside the diffusion halo and therefore $\psi_\odot$. Writing $\langle \mathcal{D}\rangle = \rho_\odot L_{\rm eff}$ for the column density of Eq.~\eqref{eq:Jgam}, the lifetime that the proton fit demands scales as $\rho_\odot \psi_\odot$ and the lifetime that the photon measurement bounds scales as $\rho_\odot L_{\rm eff}$. The local density therefore cancels identically between the two channels, and only the combination $L_{\rm eff}/\psi_\odot$ survives. We have recomputed it for nine profiles, from a Moore cusp to cores as wide as $15\kpc$, varying only the shape while keeping the geometry and the local density fixed, and the ratio stays within $+14\%$ and $-8\%$ of its NFW value. Table~\ref{tab:profiles} lists the nine individually. Two things follow. The shape of the halo is not a significant uncertainty on anything below, and its normalisation is not an uncertainty at all, which matters because the local density is the one halo parameter carrying a factor-of-two observational spread. The half-height is a separate axis, since $L_{\rm eff}$ does not depend on $H$ while $\psi_\odot$ carries the whole of that dependence, and we vary it through the propagation setups of Table~\ref{tab:prop}.

\begin{table}[t]\centering\small\setlength{\tabcolsep}{4pt}
\begin{tabular}{llccccc}\toprule
& propagation setup & $K_0$ [cm$^2$ s$^{-1}$] & $\delta$ & $H$ [kpc] & $\tau_{\rm eff}(1\PeV)$ [s] & $\lambda/H$ \\ \midrule
A & Ref.~\cite{Strong:2007nh} & $4\cdot 10^{28}$ at $3$ GV & $0.50$ & $4$ & $3.6\cdot 10^{12}$ & $0.19$ \\
B & Ref.~\cite{Genolini:2019ewc}, SLIM & $8.5\cdot 10^{28}$ at $10$ GV & $0.51 \to 0.33$ & $10$ & $4.9\cdot 10^{13}$ & $0.02$ \\
C & Ref.~\cite{Weinrich:2020ftb}, SLIM & $1.1\cdot 10^{28}$ at $1$ GV & $0.51 \to 0.33$ & $5$ & $4.4\cdot 10^{13}$ & $0.02$ \\ \midrule
LW & disc clock extrapolated \cite{Loeb:2002ee} & \multicolumn{3}{c}{$\tau = 10^{7.5}\,\mathrm{yr}\,(E/\mathrm{GeV})^{-0.6}$} & $2.5\cdot 10^{11}$ & \\
\bottomrule\end{tabular}
\caption{Effective confinement time of halo-produced nucleons at the solar position, from Eq.~\eqref{eq:Jp}. Setups B and C include the high-rigidity break at $R_h \simeq 250$ GV \cite{Genolini:2019ewc}. The last column is the ratio of the diffusion mean free path $3K/c$ to the halo half-height, quoted at $1\PeV$, which must be small for the diffusive treatment to hold. The last row is the disc-calibrated extrapolation, shown for comparison.}\label{tab:prop}\end{table}

Table~\ref{tab:prop} collects $\tau_{\rm eff}(1\PeV)$ for three published propagation setups, together with the disc-calibrated case for comparison. The spread between the three is an order of magnitude, and it is dominated by the treatment of the high-rigidity break in the diffusion coefficient, which is measured at $R_h \simeq 250$ GV with $\Delta\delta \simeq 0.18$ \cite{Genolini:2019ewc} and which flattens the energy dependence of $\tau_{\rm eff}$ from $E^{-0.5}$ to $E^{-0.33}$ over the whole range of interest.

We adopt setup B as the baseline for two reasons. It uses a complete published parameter set calibrated on AMS-02 secondary-to-primary ratios, and it gives the longest effective residence time of the three. Namely, it is the setup that demands the longest lifetime to account for the proton feature, and therefore predicts the smallest accompanying photon flux. For completeness, the analysis reported in the next section (see Table~\ref{tab:scan}) carries all four choices.

The last column of Table~\ref{tab:prop} checks that a diffusive description applies to all cases\footnote{Setup A, which has no rigidity break, reaches a range for which the diffusive approximation is no longer safely applicable. However, this should strengthen the conclusion reported in the following section. }. The ratio of the mean free path $3K/c$ to the half-height of the halo stays at $0.05$ or below over the whole range for setups B and C.

Photons, by contrast, travel in straight lines, so that
\begin{equation}\label{eq:Jgam} J_\gamma(E) \;=\; \frac{\langle \mathcal{D}\rangle}{4\pi\, m_\chi \tau_\chi}\, \frac{dN_\gamma}{dE} , \qquad \textrm{with} \qquad \langle \mathcal{D}\rangle \;=\; \frac{1}{\Delta\Omega}\int_{\Delta\Omega}\!\! d\Omega \int ds \; \rho_\chi , \end{equation}
where the column density is averaged over the observation window. For the two windows in which LHAASO measured the diffuse Galactic emission, $15^\circ < l < 125^\circ$ and $125^\circ < l < 235^\circ$ with $|b| < 5^\circ$ in both cases \cite{LHAASO:2023gne}, our NFW profile gives $\langle \mathcal{D}\rangle = 2.6 \cdot 10^{22}$ and $1.1 \cdot 10^{22} \GeV\,\mathrm{cm}^{-2}$ respectively, each over a solid angle of $0.335\,$sr. 

\section{The proton spectrum}\label{sec:fit}
We use the proton spectrum measured by LHAASO across the knee from the updated proton spectrum tabulated in \cite{LHAASO:2025mlf}, in $19$ bins of width $\Delta\log_{10}E = 0.1$ between $0.158$ and $12.6\PeV$. The measurement is published three times, once for each hadronic interaction model used to simulate the showers and to reconstruct the primary energy and composition: EPOS-LHC, QGSJETII-04, and SIBYLL 2.3d. We use all three throughout the entire analysis. The reason is that the interaction model is the largest single entry in the collaboration's systematic budget, and it is a systematic difference between three reconstructions of the same events. Therefore, it cannot be added in quadrature to the error bars of any one of them. Carrying the three tables is the only way a fit can be exposed to it. The tabulated systematic, which excludes that term, is close to flat across the range and dominates the statistical error in all but the highest bins\,\cite{LHAASO:2025mlf}.

The astrophysical background is a power law, the extrapolation of the population that dominates below $0.1\PeV$. Its index cannot be taken from the measurement we are fitting. A power law fitted to the same $19$ points would absorb part of the feature we are testing for, and would then be compared against what is left of it. We therefore take the index from an independent measurement at lower energy. DAMPE measures the proton spectrum from $40\GeV$ to $100\TeV$ and finds it softening at $13.6\TeV$, which gives $\gamma_1 = 2.85 \pm 0.07$ in the decade below the LHAASO threshold \cite{DAMPE:2019gys}, and we profile the index under that prior.

The fit has four parameters: two normalisations, the background index, and the decay mass. Only the mass is a shape parameter of the decay component. We quote $\chi^2$ against the number of points and of parameters rather than against degrees of freedom, because the prior can be counted either as a constraint on a parameter or as an additional datum. We stress that none of the $\chi^2$ in this paper is a goodness-of-fit statistic. Indeed, the fundamental take-home message from this analysis is the difference between two models evaluated under the same covariance, and not the absolute values $\chi^2$.

Moreover, the prior is not entirely independent of the hypothesis under test. Extrapolated down into the DAMPE range, the decay component supplies about a tenth of the flux at the top of it, and would therefore harden the index that DAMPE measures. Correcting for this would move the prior away from the value the LHAASO points prefer, and not towards it. Therefore, the prior as used introduces an error in the conservative direction.

The systematic is quoted bin by bin, but what a fit is sensitive to is its correlation between bins, and no covariance matrix is published.  A fully correlated systematic moves the whole spectrum up or down together, and a fit absorbs it by rescaling the normalisation, so it constrains nothing about shape. An uncorrelated systematic is point-to-point scatter, which a fit can absorb only bin by bin. The feature we are testing is a shape, so how much of the systematic is correlated decides how much of that shape is really there.

Treated as uncorrelated, the best simple astrophysical description, a power law with an exponential cutoff, returns about one per degree of freedom in each reconstruction. This does not validate the error model, but it rules out the failure that would matter here, an error model so generous that any smooth curve passes through it. Treated as fully correlated, the systematic leaves only the statistical errors, which are of the order of one part in a thousand in the lowest bins. No smooth function follows $19$ points at that precision, and neither the astrophysical reference nor the decay model survives that limit, both returning $\chi^2$ in the hundreds. The measurement, therefore, excludes the fully correlated case and accommodates the uncorrelated one. We use a diagonal covariance throughout and carry an intermediate family as a check; details are reported in Appendix~\ref{app:cov} where we also explain why a single parameter cannot be calibrated to one value.

\begin{figure}[t]\centering \includegraphics[width=\textwidth]{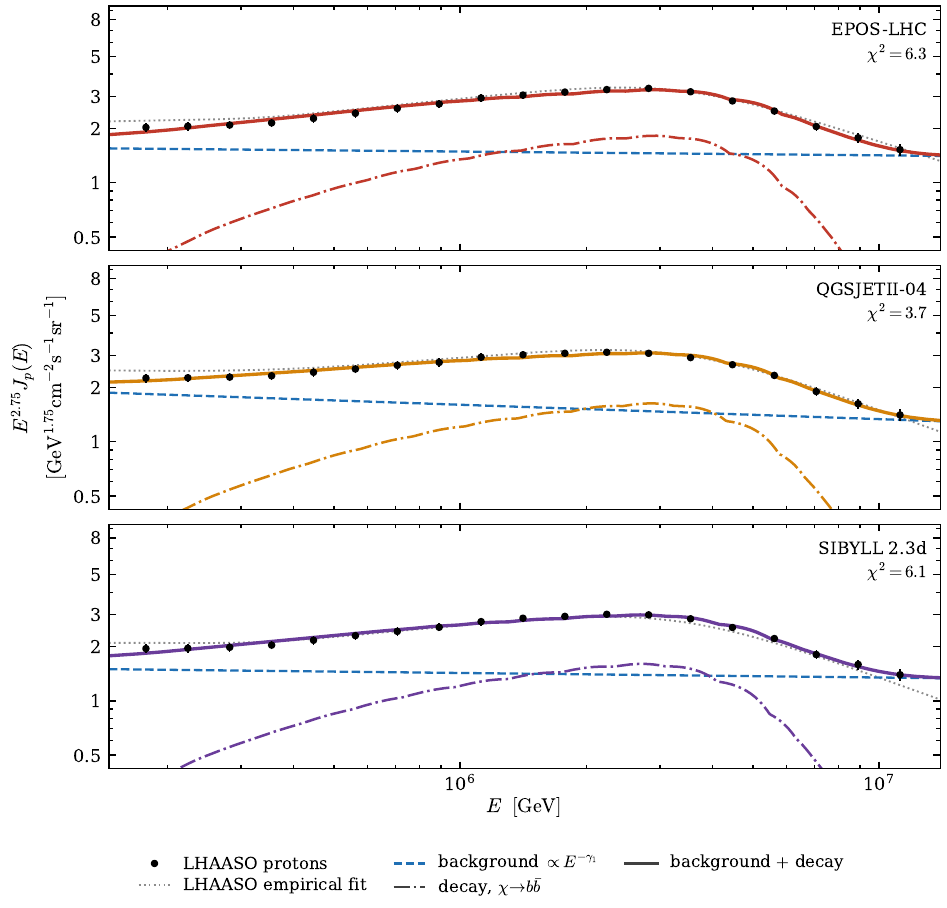}
\caption{The LHAASO proton spectrum in the three hadronic-interaction-model reconstructions, fitted with a power law plus the nucleons from the decay of a superheavy relic. The decay contribution is dash-dotted, and the dotted curve is the collaboration's own empirical fit \cite{LHAASO:2025byy}. }\label{fig:spectrum}\end{figure}

Adding the decay contribution of Eq.~\eqref{eq:Jp} to that background and profiling over the normalisations at each mass, we obtain
\begin{equation}\label{eq:bestfit} \begin{array}{lcccc} \text{EPOS-LHC:} & m_\chi = 4.9\cdot 10^{7}\GeV , & \tau_\chi = 2.5\cdot 10^{25}\,\mathrm{s} , & \gamma_1 = 2.77 , & \chi^2 = 6.3 , \\ \text{QGSJETII-04:} & m_\chi = 4.9\cdot 10^{7}\GeV , & \tau_\chi = 2.8\cdot 10^{25}\,\mathrm{s} , & \gamma_1 = 2.83 , & \chi^2 = 3.7 , \\ \text{SIBYLL 2.3d:} & m_\chi = 4.6\cdot 10^{7}\GeV , & \tau_\chi = 2.8\cdot 10^{25}\,\mathrm{s} , & \gamma_1 = 2.78 , & \chi^2 = 6.1 , \end{array} \end{equation}
in each case for $19$ points and four fitted parameters, and shown in Fig.~\ref{fig:spectrum}. The description is equally good in all three reconstructions. The fitted masses span $7\%$ and the lifetimes $11\%$, although the three input spectra differ by up to $12\%$ in normalisation and by $0.05$ in effective slope. The shape that the fit selects is therefore common to all of them.

The nucleon spectrum from a decay turns over at an energy well below $m_\chi/2$. For the fitted mass \texttt{HDMSpectra} puts that turnover at $x = 0.13$, which is $3.3\PeV$, and it is this feature, and not the endpoint at $x = 1$, that the fit matches to the knee. The older analytic form of Eq.~\eqref{eq:gauss} has no such turnover. It runs on to $x = 1$ and is then cut off by hand. Fitted to the same data, the only feature it can offer is that artificial edge, so it places $m_\chi/2$ at the observed knee, where the full calculation places it nearly eight times higher.

Moreover, the result does not depend on the decay channel. Repeating the fit for $u\bar u$, $t \bar t$, $gg$, $W^+W^-$, and $hh$, the lifetime is stable to within a factor of $1.6$ while the mass moves by a factor of $3.6$. The mass is the less robust of the two because it is fixed by the position of the fragmentation fall-off, which the channel controls, whereas the lifetime is fixed by the flux the excess has to supply.

The fit above shows that a decaying relic can describe the feature. Whether the data require one is a separate question, which can only be answered against an astrophysical alternative, and the answer depends on which alternative we choose. The selected alternatives are reported in Table.\,\ref{tab:bkg} and summarised below.

\begin{table}[bth!]\centering\small
\begin{tabular}{lccccc}\toprule
astrophysical background & par. & dof & $\chi^2$ alone & $\chi^2$ with decay & $\Delta\chi^2$ \\ \midrule
power law with a cutoff            & $3$ & $16 \to 14$ & $18.3$ / $13.2$ / $18.1$ & $4.9$ / $3.6$ / $4.8$ & $13.4$ / $9.6$ / $13.3$ \\
two populations, soft index fixed  & $4$ & $15 \to 13$ & $13.5$ / $11.8$ / $12.5$ & $1.5$ / $2.8$ / $1.5$ & $12.0$ / $9.0$ / $11.1$ \\
two populations, soft index free   & $5$ & $14 \to 12$ & $1.1$ / $0.5$ / $0.9$ & $0.7$ / $0.3$ / $0.6$ & $0.4$ / $0.2$ / $0.3$ \\
collaboration function, refitted   & $8$ & $11 \to 9$ & $0.8$ / $0.8$ / $1.1$ & $0.4$ / $0.2$ / $0.3$ & $0.4$ / $0.6$ / $0.9$ \\
\bottomrule\end{tabular}
\caption{Four astrophysical backgrounds, each fitted alone and with the decay term of Eq.~\eqref{eq:Jp} added. Entries are EPOS-LHC / QGSJETII-04 / SIBYLL 2.3d.}\label{tab:bkg}\end{table}

The first row of Table.\,\ref{tab:bkg} is based on a power law with a cutoff. Adding the decay term to it improves $\chi^2$ by between $10$ and $13$, and by between $9$ and $14$ whichever covariance we use. Neither of these two fits carries the DAMPE prior, so their $\chi^2$ are not those of Eq.~\eqref{eq:bestfit}. The improvement looks like strong support, but the test is unfair. The data harden before they steepen, and a power law with a cutoff can only soften. The comparison model is not allowed to produce a hardening at all, so part of what the decay is credited with is simply the fact that it can.

We therefore repeat the comparison against backgrounds that can harden. Two populations, a soft one and a harder one with a cutoff, are the simplest, and they are the second and third rows of Table~\ref{tab:bkg}. With the soft index held at the published value, the reference sits at about $0.85$ per degree of freedom, and the decay still brings it down to $0.1$ to $0.2$, so the improvement survives. If we release that index, however, so that the astrophysical model has five free parameters, exactly as many as the decay model, it already describes the spectrum at below $0.1$ per degree of freedom on its own, and adding the decay leaves it below $0.1$. There is nothing left for the decay to improve. The collaboration's own eight-parameter function, in the last row, behaves the same way.

Those numbers say something about the flexible backgrounds as well as about the decay. A model that reaches $0.1$ per degree of freedom is not describing the data; it is describing the error bars, and both of the last two rows have enough freedom to do so. That is the sense in which the comparison stops being informative: at equal complexity, the two descriptions are indistinguishable, and at greater complexity, neither is constrained at all. 

For completeness, we have repeated the identical fit with the Gaussian fragmentation spectrum of Eq.~\eqref{eq:gauss}. It returns a parent mass lower by a factor of eight to nine, and a $\chi^2$ an order of magnitude worse. Again, the reason for this difference is that the form does not have an endpoint.

\section{The photon counterpart}\label{sec:gamma}
Whatever produces PeV nucleons through a parton cascade produces PeV photons at the same time, and Sec.~\ref{sec:setup} showed that the photons carry the larger share of the energy. LHAASO has measured the diffuse Galactic emission at exactly these energies \cite{LHAASO:2023gne}. In this section, we use that measurement to rule out the decay.

\subsection{Comparison with the diffuse Galactic emission}
LHAASO measures the diffuse emission in two windows along the Galactic plane, one inner and one outer. The LHAASO analysis blanks out the regions around bright sources and compares models only to the sky that is left, so we use the published mask and do the same. It removes about a third of the inner window, but it changes the mean halo column by only a few per cent, because the blanked regions are small and scattered instead of being concentrated where the halo is brightest.

Absorption has to be included as well. A PeV photon can pair-produce on the cosmic microwave background before it reaches us. We compute that loss directly, and as a check, our attenuation length at $2.2\PeV$ agrees with the published one \cite{Vernetto:2016alq}. The attenuation is negligible below approximately $100\TeV$ and removes about half the flux in the highest bin. The interstellar radiation field absorbs as well, but we do not calculate it. For simplicity, we apply the largest absorption probability quoted anywhere in Ref.~\cite{Vernetto:2016alq}, which belongs to lines of sight through the Galactic centre that our windows do not contain. That over-corrects, and since absorption only lowers our predicted flux, it should be taken as a conservative estimate.

\begin{figure}[t]\centering \includegraphics[width=\textwidth]{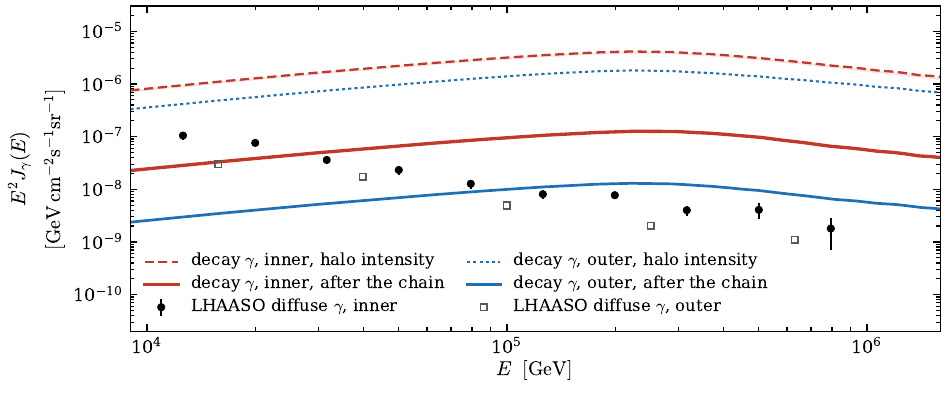}
\caption{The photon flux produced by the same decay that Fig.~\ref{fig:spectrum} computes for the proton feature, compared with the LHAASO diffuse Galactic emission in the inner and outer plane windows \cite{LHAASO:2023gne}. Dashed lines are the halo intensity and solid lines are the same prediction passed through the two-step background chain of Appendix~\ref{app:onoff}. Each line shows the three hadronic models that almost overlap, making it impossible to distinguish. }\label{fig:gamma}\end{figure}

With both included, the predicted flux exceeds the measured one in every bin, by about an order of magnitude at the bottom of the range and by three orders at the top. The gap widens because the two spectra have different shapes: the measured emission falls steeply with energy, whereas the decay spectrum stays nearly flat in $E^2dN/dE$ until the fragmentation fall-off sets in. 
The three reconstructions give almost the same prediction, and in Fig.~\ref{fig:gamma} the band of three lines appears as a single line\footnote{Our curves are evaluated at the energy quoted for each bin, whereas the measurement integrates over the bin and folds in the detector resolution. Doing the same to our prediction moves it by a few per cent, and upward in the highest bins, where the overshoot is already largest. The comparison as we make it is therefore slightly conservative.}.

\subsection{Background subtraction chain}\label{sec:bkgr subtraction}
We do not need a model of the astrophysical $\gamma$-ray emission, since we ask only whether the decay alone already exceeds everything that was measured. What we do need is how LHAASO estimated its own cosmic-ray background.

That estimate is taken from the sky next to the plane, at the same declination but other right ascensions, with the plane itself blanked out. For emission that lies in the plane this works well, because the region used for the background holds almost none of the signal. A halo signal is not confined to the plane: it fills the sky, so it sits in the background region as well and is subtracted from the measurement along with the background. The analysis then smooths the count maps on a $20^\circ$ scale to remove whatever large-scale structure is left, and a halo component is large-scale structure. Both steps are reported in Appendix~\ref{app:onoff}. We find that the first step leaves about a sixth of the raw halo column in the inner window, and that the second takes it down to a few per cent. The consequence is two numbers rather than one: on the raw flux the decay is excluded by a factor of several hundred, which is what an analysis keeping the signal in both regions would constrain, as a dedicated LHAASO dark matter search does \cite{LHAASO:2022yxw}, whereas after the chain the factor is about twenty, which is what this measurement, as published, can exclude.

The factor of twenty depends on our reconstruction of LHAASO's background procedure, making this the most uncertain step in the analysis so far. 
In order to reduce this uncertainty, we can refer to the following comparison that requires no background model, no target region, and no subtraction. Photon-induced air showers contain far fewer muons than proton-induced ones because they develop electromagnetically. KASCADE uses this to bound the fraction of the cosmic radiation above a given shower size that is photons \cite{KASCADEGrande:2017vwf}. To compare, we integrate our predicted photon flux from $3.7\PeV$ up to the endpoint at $m_\chi/2$ and weight it by the exposure of an array at latitude $49.1^\circ$N looking within $20^\circ$ of the zenith. The weighting matters because such an array never sees the Galactic centre, and therefore never sees the brightest part of the halo. Varying the uncertainty in this computation, with more details in Appendix~\ref{app:kascade}, we find that the exclusion remains between $60-900$, thus, our claim remains robust.

\subsection{The exclusion and the origin of the relic}
The analysis above adopts a single propagation setup. For a fixed mass and decay spectrum, the proton flux scales as $\tau_{\rm eff}(E)/\tau_\chi$. Longer confinement therefore allows the same proton flux to be produced with a lower decay rate, corresponding to a longer lifetime. Since the photon flux scales as $1/\tau_\chi$ and is unaffected by magnetic confinement, this reduces the photon signal associated with the proton feature. An overall change in $\tau_{\rm eff}$ can be absorbed into the fitted lifetime, but a change in its energy dependence also modifies the spectral shape. We therefore repeat the full analysis for each propagation setup and each hadronic-interaction-model reconstruction, with the results reported in Table~\ref{tab:scan}. All twelve combinations remain excluded by the photon comparison, with exclusion factors of approximately $20$ to $5000$ after the background-subtraction procedure.

The origin of the relic abundance is a separate question. The fitted mass, $m_\chi\simeq5\cdot10^7\GeV$, lies more than two orders of magnitude above the unitarity bound for a conventional thermal freeze-out relic \cite{Griest:1989wd}. Obtaining the observed dark matter abundance therefore requires a different production mechanism or a departure from the standard freeze-out history. Gravitational production during the transition out of inflation is one possibility. For the production history considered in Ref.~\cite{Chung:2001cb}, and in the regime $m_\chi<H_I$, the abundance is estimated as
\begin{equation}
\Omega_\chi h^2 \simeq
\left(\frac{m_\chi}{10^{11}\GeV}\right)^2
\left(\frac{T_{\rm RH}}{10^9\GeV}\right),
\end{equation}
where $H_I$ is the Hubble expansion rate at the end of inflation. Substituting the fitted mass and $\Omega_\chi h^2\simeq0.12$ gives $T_{\rm RH}\simeq5\cdot10^{14}\GeV$. This is a formal estimate within the quoted scaling relation; establishing a viable production scenario also requires checking that the assumed inflationary and reheating history applies at these parameters.

Our exclusion does not require a particular production mechanism. Assuming that the relic constitutes all of the dark matter, the lifetime needed to reproduce the proton feature is shorter than the photon observations allow. More generally, both signals scale with the relic density divided by its lifetime. Reducing the relic abundance therefore does not remove the photon excess if the lifetime is adjusted to preserve the proton contribution, provided the relic traces the same halo distribution. The excluded hypothesis is that these decays explain the proton feature; a relic of the same mass with a sufficiently lower decay rate remains possible\footnote{Related constraints on specific gravitationally produced dark matter candidates have been derived from LHAASO diffuse emission combined with Fermi-LAT data \cite{Barman:2025gjr}.}.

\begin{table}[t]\centering\small
\begin{tabular}{lcccccc}\toprule
setup & $\tau_{\rm eff}(1\PeV)$ [s] & $m_\chi$ [GeV] & $\tau_\chi$ [s] & $\chi^2$ (19 pts, 4 par.) & \multicolumn{2}{c}{exclusion factor} \\
\cmidrule(lr){6-7}
 & & & & & raw flux & after the chain \\ \midrule
\multicolumn{7}{l}{\it EPOS-LHC} \\
A  & $3.6\cdot 10^{12}$ & $5.9\cdot 10^{7}$ & $1.6\cdot 10^{24}$ & $16.5$ & $1.0\cdot 10^{4}$ & $3.3\cdot 10^{2}$ \\
B  & $4.9\cdot 10^{13}$ & $4.9\cdot 10^{7}$ & $2.5\cdot 10^{25}$ & $6.3$ & $6.8\cdot 10^{2}$ & $2.1\cdot 10^{1}$ \\
C  & $4.4\cdot 10^{13}$ & $4.9\cdot 10^{7}$ & $2.3\cdot 10^{25}$ & $6.2$ & $7.5\cdot 10^{2}$ & $2.4\cdot 10^{1}$ \\
LW & $2.5\cdot 10^{11}$ & $6.4\cdot 10^{7}$ & $9.9\cdot 10^{22}$ & $25.3$ & $1.6\cdot 10^{5}$ & $5.1\cdot 10^{3}$ \\
\midrule
\multicolumn{7}{l}{\it QGSJETII-04} \\
A  & $3.6\cdot 10^{12}$ & $5.3\cdot 10^{7}$ & $1.7\cdot 10^{24}$ & $7.6$ & $9.7\cdot 10^{3}$ & $3.1\cdot 10^{2}$ \\
B  & $4.9\cdot 10^{13}$ & $4.9\cdot 10^{7}$ & $2.8\cdot 10^{25}$ & $3.7$ & $6.1\cdot 10^{2}$ & $1.9\cdot 10^{1}$ \\
C  & $4.4\cdot 10^{13}$ & $4.9\cdot 10^{7}$ & $2.5\cdot 10^{25}$ & $3.8$ & $6.8\cdot 10^{2}$ & $2.1\cdot 10^{1}$ \\
LW & $2.5\cdot 10^{11}$ & $6.0\cdot 10^{7}$ & $1.1\cdot 10^{23}$ & $13.2$ & $1.5\cdot 10^{5}$ & $4.7\cdot 10^{3}$ \\
\midrule
\multicolumn{7}{l}{\it SIBYLL 2.3d} \\
A  & $3.6\cdot 10^{12}$ & $5.2\cdot 10^{7}$ & $1.8\cdot 10^{24}$ & $13.3$ & $9.7\cdot 10^{3}$ & $3.0\cdot 10^{2}$ \\
B  & $4.9\cdot 10^{13}$ & $4.6\cdot 10^{7}$ & $2.8\cdot 10^{25}$ & $6.1$ & $6.2\cdot 10^{2}$ & $1.9\cdot 10^{1}$ \\
C  & $4.4\cdot 10^{13}$ & $4.6\cdot 10^{7}$ & $2.5\cdot 10^{25}$ & $6.0$ & $6.9\cdot 10^{2}$ & $2.2\cdot 10^{1}$ \\
LW & $2.5\cdot 10^{11}$ & $6.0\cdot 10^{7}$ & $1.1\cdot 10^{23}$ & $21.3$ & $1.5\cdot 10^{5}$ & $4.6\cdot 10^{3}$ \\
\bottomrule\end{tabular}
\caption{The analysis repeated for the four propagation setups of Table~\ref{tab:prop} and all three reconstructions. The exclusion factor is the ratio of the $\gamma$-ray bound at the fitted mass to the fitted lifetime, quoted for the raw flux and after the background chain.}\label{tab:scan}\end{table}

\section{Bounds from the two channels}\label{sec:bound}
Both measurements can be used to set limits on the relic lifetime. As a simple procedure, we require that the predicted flux not exceed the measured flux by more than twice its error in any bin, and we apply that rule to the $19$ proton bins of Ref.~\cite{LHAASO:2025mlf} and to the $10$ inner-window photon bins of Ref.~\cite{LHAASO:2023gne}, the latter with the mask and the absorption of Sec.~\ref{sec:gamma}.

This construction is deliberately conservative, and it is not a confidence limit. Indeed, we simply ask that the prediction stay below every measured point by a fixed margin. Its coverage over bins whose systematic errors are certainly not independent is not the nominal one, we have not calibrated it, and we therefore attach no confidence level to either curve\footnote{The requirement that a putative signal not overshoot the observed flux in any energy bin, with no background subtracted, is done in Ref.~\cite{Ishiwata:2019aet}, which applies it to the cosmic-ray proton spectrum measured by the Pierre Auger Observatory above $10^{9}\GeV$, and to positron and neutrino data; where that work uses a measured diffuse $\gamma$-ray flux instead, it models the foreground and tests against it. It saturates the measured central value, whereas we let the prediction reach the measurement plus twice its error, a tolerance that makes both of our bounds weaker than the same rule at saturation.}.

\begin{figure}[t]\centering \includegraphics[width=\textwidth]{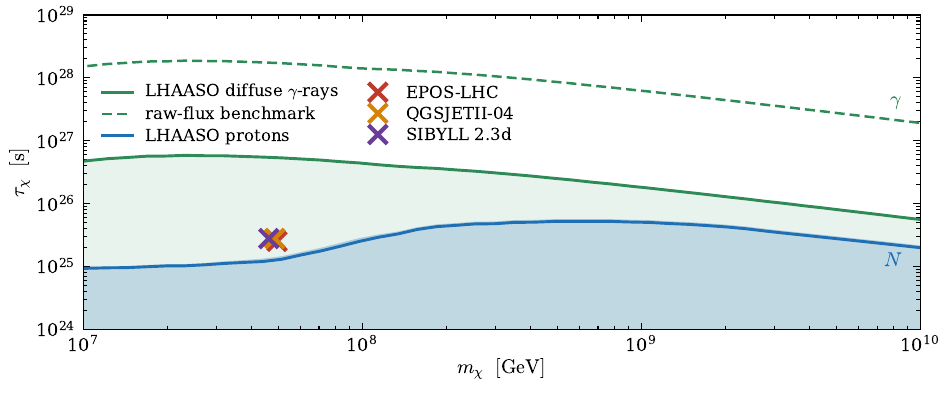}
\caption{Binwise no-overshoot benchmarks for the two channels. The blue band is derived from the LHAASO proton spectrum \cite{LHAASO:2025mlf} and spans the three hadronic-interaction-model reconstructions. The green curves are derived from the LHAASO diffuse Galactic emission \cite{LHAASO:2023gne}: the solid curve includes the background-subtraction chain of Sec.~\ref{sec:bkgr subtraction}, while the dashed curve is the corresponding benchmark on the raw halo intensity. Crosses mark the proton best-fit parameters of Sec.~\ref{sec:fit}.}\label{fig:plane}\end{figure}

The nucleon bound is strongest near $m_\chi \simeq 7\cdot 10^{8}\GeV$, where the energy-weighted peak of the fragmentation spectrum falls inside the LHAASO window, and reads
$\tau_\chi \gtrsim 5 \cdot 10^{25}\,\mathrm{s}\ $
for all the configurations. It is the one quantity in this paper that is insensitive both to the choice of reconstruction and to the background model, because it is set by the measured flux and its error rather than by the shape of any feature in it. It constrains decaying superheavy dark matter from a direct measurement of the cosmic-ray nucleon flux, lying inside the $10^{2}$ to $10^{9}\GeV$ interval that Ref.~\cite{Ishiwata:2019aet} had to leave empty.

The two results of the bound of this paper rely on different parts of  the fragmentation calculation. The fit of Sec.~\ref{sec:fit} probes $x$ between about $10^{-2}$ and $0.5$, comfortably inside the range where Ref.~\cite{Bauer:2020jay} quotes ten per cent accuracy. The bound at the mass where it peaks reaches down to $x \sim 10^{-3}$ and below, where that reference reports uncertainties of order unity from an incomplete treatment of coherent soft emission. The high-mass end of the bound therefore carries an uncertainty that the fit does not.

The photon bound obtained with the same criterion is stronger than the nucleon one. It is useful to locate it against the literature, since it comes from a single measurement with no likelihood and no modelling of the astrophysical background, and should therefore be somewhat weaker than a dedicated analysis. This is consistent with the published results: the combined HAWC and LHAASO analysis of Ref.~\cite{Rocamora:2025ddt} reaches almost $10^{29}\,$s at the top of its mass range, and Ref.~\cite{Dubey:2025ouh} excludes lifetimes comparable to our raw-flux benchmark at $10^{7}\GeV$. Being a factor of a few below dedicated analyses is the expected behaviour of a no-overshoot criterion applied to one dataset, and that agreement checks the whole calculation, the halo profile and column densities of Sec.~\ref{sec:setup} as much as the fragmentation spectra.

Because both bounds use measurements from the same experiment and the same criterion, their ratio shows how much more strongly one channel constrains the lifetime. Over the mass range shown in Fig.~\ref{fig:plane}, the photon bound is one to two orders of magnitude stronger than the nucleon bound, or two to three orders of magnitude stronger when using the raw-flux benchmark. This comparison builds on the point made in Sec.~\ref{sec:setup}: the decay produces a substantial photon component alongside the nucleons. The relative strength of the bounds depends on this energy budget, the halo geometry discussed in Sec.~\ref{sec:prop}, and the much larger background against which the nucleon signal must be detected. Using a confinement time calibrated on disc sources would underestimate the nucleon signal and make its bound a further two orders of magnitude weaker relative to the photon bound.

\section{Conclusions}\label{sec:concl}
By resolving the proton component through the knee, LHAASO makes it possible to ask whether the observed structure could be produced by nucleons from a superheavy relic decaying in the Galactic halo. When the background index is profiled under an independent prior on the proton slope below $0.1\PeV$, a relic with mass $5\cdot 10^{7}\GeV$ and lifetime of a few times $10^{25}\mathrm{s}$ reproduces both the hardening and the knee, with its mass as the only shape parameter. The fit is equally good in all three hadronic-interaction-model reconstructions. We do not, however, interpret the quality of the fit as evidence for the model. The covariance of the dominant systematic uncertainty is not published, the improvement disappears when the comparison is made against a two-population background with the same number of parameters, and the inferred mass varies by a factor of $3.6$ across the decay channels considered.

The same cascade also produces photons, and this counterpart rules out the interpretation. Hadronisation transfers about three times as much energy to photons as to nucleons, almost independently of the decay channel and parent mass. Even after the prediction is passed through the two-stage background procedure used in the LHAASO diffuse analysis, which removes most of the extended halo component, the lower limit on the lifetime remains about twenty times larger than the lifetime required by the proton fit. The KASCADE-Grande limit on the photon fraction above $3.7\PeV$ provides an independent test that requires neither a background model nor a subtraction procedure. Its nominal exclusion factor is about $200$; varying the assumptions entering the recast broadens this range to approximately $60$--$900$. Both exclusions persist for every propagation setup, hadronic-interaction-model reconstruction, and decay channel that we have examined.

Source geometry is essential for the nucleon calculation. Per unit local source rate, a source filling the halo maintains a local density about $35$ times larger than one confined to the thin disc for which radioactive clocks calibrate the residence time. Solving the diffusion problem for the halo geometry turns the proton measurement into a direct bound, $\tau_\chi \gtrsim 5\cdot 10^{25}\mathrm{s}$ near $m_\chi \simeq 7\cdot 10^{8}\GeV$, in the mass range left uncovered by previous hadronic studies. Second, the Gaussian SUSY-QCD spectrum commonly used in the top-down literature has no endpoint suppression and would return a parent mass nearly an order of magnitude lower. Searches for a localised feature from a decaying relic must therefore use a modern fragmentation calculation. A dedicated anisotropy analysis of the LHAASO nucleon sample could test the predicted dipole towards the Galactic centre, while publication of the proton covariance matrix would determine whether the spectral resemblance carries any statistical significance.

\section*{Acknowledgements}
D.A., D.K. and A.V.M. acknowledge support from Tamkeen under the NYU Abu Dhabi Research Institute grant CASS.
A.J.I.'s research is funded by Tamkeen under the NYU Abu Dhabi Research Institute grant ADHPG-AD457.

\appendix

\section{The diffusion calculation}\label{app:diffusion}
Equation~\eqref{eq:Jp} follows from solving $-K\nabla^2 n = Q$ in a cylinder of radius $R = 20\kpc$ and half-height $H$ centred on the Galactic centre, with free escape at both boundaries and $Q \propto \rho_\chi$. It reduces to Poisson's equation for the normalised field $\psi$ defined by $-\nabla^2\psi = \rho_\chi/\rho_\odot$.

We stress that the $\tau_{\rm eff}$ this produces is a steady-state response, namely the density that a given source rate maintains at the Sun, divided by that rate. It has the dimensions of time, but it is not the mean age of the particles. The mean age is a property of their arrival-time distribution, and it is what a radioactive clock measures. The two coincide only for a source geometry that the clock and the halo do not share, which is why the disc-calibrated residence time cannot simply be imported.

The source is the halo profile truncated at the boundaries, and nothing beyond it. This is the standard treatment: the boundary condition models rapid escape, and within that model, a particle produced outside does not return. That is an assumption about the boundary and not a fact about the halo. A more permeable boundary would raise $\psi_\odot$ and therefore $\tau_{\rm eff}$, which lengthens the lifetime the proton fit requires and weakens the photon exclusion rather than strengthening it, so the assumption makes an error in the conservative direction. The photon column density of Eq.~\eqref{eq:Jgam} has no such boundary and is integrated out to the virial radius.

The equation is discretised in cylindrical coordinates on a uniform grid and solved directly as a sparse linear system. Doubling the resolution changes $\psi_\odot$ by less than half a per cent, and for a uniform source in a wide slab, the solution reproduces the analytic result $\psi_\odot = H^2/2$ to four digits, which validates the scheme. 
The diffusion coefficient is taken to be spatially uniform, as in the frameworks the parameter sets come from, and is evaluated at the rigidity of the particle; for protons and antiprotons, the only species entering here, rigidity and energy coincide numerically in the units used.

The halo-profiles reported in Sec.~\ref{sec:prop} are run through the same solver at fixed geometry and results are reported in Table~\ref{tab:profiles}.  The stability of the ratio $L_{\rm eff}/\psi_\odot$ has a simple origin. Both quantities are dominated by the region from the solar circle outwards, where every profile is pinned to the same measured local density, and neither is sensitive to the innermost few kiloparsecs. The diffusion cylinder is centred on the Galactic centre, but what enters is Eq.~\eqref{eq:Jp} evaluated at the solar position, and the Green function of the diffusion operator weights the source by its distance from that point. The cusp is therefore not what determines $\psi_\odot$, and the $\gamma$-ray windows are independent of the Galactic centre in any case. Repeating the entire set of profiles at the smallest half-height among our setups widens the spread only slightly, so the insensitivity to shape is not a property of the baseline height.

\begin{table}[t]\centering\small\setlength{\tabcolsep}{4pt}
\begin{tabular}{lccccrr}\toprule
& \multicolumn{2}{c}{$\psi_\odot$ [kpc$^2$]} & & & \multicolumn{2}{c}{$L_{\rm eff}/\psi_\odot$ vs NFW [\%]} \\
\cmidrule(lr){2-3} \cmidrule(lr){6-7}
profile & $H = 10\kpc$ & $H = 4\kpc$ & $L_{\rm eff}$ [kpc] & $L_{\rm eff}/\psi_\odot$ & $H = 10$ & $H = 4$ \\ \midrule
NFW, $r_s = 20\kpc$ (baseline) & $34.72$ & $8.70$ & $21.52$ & $0.620$ & $0$ & $0$ \\
NFW, $r_s = 10\kpc$ & $36.90$ & $9.33$ & $20.99$ & $0.569$ & $-8.2$ & $-9.0$ \\
Einasto, $\alpha = 0.17$ & $36.35$ & $9.13$ & $21.98$ & $0.605$ & $-2.4$ & $-2.6$ \\
Moore cusp & $35.35$ & $8.80$ & $22.14$ & $0.626$ & $+1.1$ & $+1.7$ \\
Burkert core, $r_0 = 5\kpc$ & $33.79$ & $8.71$ & $19.46$ & $0.576$ & $-7.1$ & $-9.7$ \\
Burkert core, $r_0 = 9.26\kpc$ & $30.31$ & $7.67$ & $18.98$ & $0.626$ & $+1.0$ & $+0.0$ \\
Burkert core, $r_0 = 15\kpc$ & $30.70$ & $7.49$ & $21.70$ & $0.707$ & $+14.0$ & $+17.2$ \\
cored isothermal, $r_c = 2\kpc$ & $36.66$ & $9.25$ & $22.67$ & $0.618$ & $-0.2$ & $-1.0$ \\
cored isothermal, $r_c = 5\kpc$ & $31.51$ & $7.93$ & $21.48$ & $0.682$ & $+10.0$ & $+9.5$ \\
\bottomrule\end{tabular}
\caption{The nine halo profiles behind the robustness statement of Sec.~\ref{sec:prop}, at fixed geometry and common local density. The exclusion factor is proportional to $L_{\rm eff}/\psi_\odot$, so the last two columns are the whole of the halo-shape uncertainty on it. }\label{tab:profiles}\end{table}

Finally, the parameter sets of Table~\ref{tab:prop} are not independent measurements of $K_0$ and $H$. Secondary-to-primary ratios constrain the combination $K_0/H$, so varying the two independently overstates the uncertainty on anything built from them. Along the direction the data actually fix, $\tau_{\rm eff} = \psi_\odot(H)/K_0$ with $K_0 \propto H$, and since $\psi_\odot$ grows more slowly than $H^2$ for a halo source, the residual dependence is weak. Over the whole plausible range of halo heights, it amounts to a factor of about three, against a factor of fourteen between the published setups and two orders of magnitude once the disc extrapolation is included. The halo height is therefore not the dominant uncertainty in the propagation, unlike the treatment of the high-rigidity break in the diffusion coefficient.

\section{The error model}\label{app:cov}

No covariance matrix is published for the proton spectrum. We therefore use a diagonal covariance in Sec.\,\ref{sec:fit} and test how correlations between bins affect the result.

The tabulated systematic excludes the hadronic-interaction-model dependence, which we test through the three reconstructions, and the energy-scale uncertainty, which is a separate contribution. What remains is dominated by the composition assumption. This uncertainty is expected to be correlated across bins, since an incorrect composition biases neighbouring bins in the same direction, but the published errors do not specify the strength or extent of that correlation.

We test a family of covariance matrices that interpolates between uncorrelated and fully correlated systematic errors,
\begin{equation}
C_{ij}
=
\left(\sigma_i^{\rm stat}\right)^2\delta_{ij}
+
\sigma_i^{\rm syst}\sigma_j^{\rm syst}
\exp\!\left[-\frac{|\log_{10}(E_i/E_j)|}{\ell}\right],
\end{equation}
where \(\ell\) is the correlation length in \(\log_{10}E\). The limit \(\ell\to0\) gives the diagonal covariance used in the main analysis, while \(\ell\to\infty\) makes the systematic fully correlated.

One possible way to choose \(\ell\) is to require the astrophysical reference model to return \(\chi^2\) per degree of freedom equal to one. This prescription does not give a unique answer. The reference \(\chi^2\) is not monotonic in \(\ell\): correlations over roughly a bin width can accommodate structure in the residuals, whereas correlations over much larger scales cannot. The curve reaches a minimum near \(\ell=0.12\) before rising steeply towards the fully correlated limit. It crosses unity twice for two reconstructions and once for the third. This criterion, therefore, cannot determine a single correlation length, nor would it establish the true covariance.

We retain the diagonal covariance as the baseline and use this family as a sensitivity check. Between \(\ell=0\) and \(\ell=0.1\), the fitted mass changes by only a few per cent. Across the covariance choices tested in Sec. 3, adding the decay contribution to a power law with a cutoff improves \(\chi^2\) by about \(9\)–\(14\). These checks support the stability of the reported comparison, but do not determine the correlations in the measurement. A published covariance matrix, or a decomposition of the systematic into nuisance parameters, would allow those correlations to be treated directly.

\section{Removing the halo component}\label{app:onoff}
The LHAASO collaboration~\cite{LHAASO:2023gne} estimates its cosmic-ray background in two steps, and both remove a smooth halo component.

The first is direct integration. At each time step, the event distribution in detector coordinates is built from the events arriving within $\pm 5$ hours, which at fixed declination is an average over $\pm 75^\circ$ of right ascension, with the Galactic plane removed from that average and the source mask applied throughout. What the analysis measures in a given direction is therefore not the intensity there, but the difference between it and that average. For an emission confined to the plane, the two are similar, since the background region carries almost none of the signal. For a halo component, they are not, and the residual is
\begin{equation}\label{eq:stage1} \mathcal{R}(\alpha,\delta) \;=\; \mathcal{D}(\alpha,\delta) \;-\; \big\langle \mathcal{D}(\alpha',\delta) \big\rangle_{|\alpha'-\alpha| < 75^\circ} , \end{equation}
where the average runs over the unmasked right ascensions only.

The second step is the correction for large-scale structure. The long time window leaves spurious structure in the background, and the analysis removes it by smoothing the masked ON and OFF count maps with a top-hat kernel of $20^\circ$ radius and multiplying the estimated background, pixel by pixel, by the ratio of the two smoothed maps. Writing the ON map as $B + S$ and the direct-integration background as $B + \langle S\rangle$, with $B$ the cosmic-ray background and $S$ the halo signal, the correction adds $\langle\mathcal{R}\rangle_{20^\circ}$ to the background and $B$ drops out, so what survives the whole chain is
\begin{equation}\label{eq:stage2} \mathcal{T}(\alpha,\delta) \;=\; \mathcal{R}(\alpha,\delta) \;-\; \big\langle \mathcal{R} \big\rangle_{20^\circ} . \end{equation}
This is an operation designed to remove structure on $20^\circ$ scales, and a halo component is exactly such a structure.

We evaluate both steps on a grid covering the KM2A field of view, with the absorption of Sec.~\ref{sec:gamma} applied throughout, and average over each window. Direct integration alone leaves about a sixth of the raw column in the inner window and a deficit in the outer one. The second step takes the surviving fraction down to $0.030$ and $0.007$. Both steps depend only weakly on energy, because path-length effects largely cancel in the ratio between the window and the comparison regions.

\section{The KASCADE-Grande recast}\label{app:kascade}
The KASCADE collaboration\,\cite{KASCADEGrande:2017vwf} limits the ratio of integral fluxes $I_\gamma(>E)/I_{\rm CR}(>E)$, obtained by selecting showers above a threshold in size and converting that threshold into an energy under an assumed spectral index. We integrate the predicted photon flux from the quoted energy up to the endpoint at $m_\chi/2$, and the all-particle flux over the same range from the broken power law measured by LHAASO-KM2A \cite{LHAASO:2024knt}.

The published limit assumes an isotropic flux, and a halo signal is not isotropic over the sky the array observes. For an array at geographic latitude $\phi$ observing at a uniform rate with a zenith cut $\theta_{\rm max}$, the relative exposure at declination $\delta$ is $\cos\phi\cos\delta\sin\alpha_m + \alpha_m\sin\phi\sin\delta$, where $\alpha_m$ is the hour angle at which the cut is reached, $\cos\alpha_m = (\cos\theta_{\rm max} - \sin\phi\sin\delta)/(\cos\phi\cos\delta)$, taken as $0$ or $\pi$ when that expression leaves the unit interval. At the latitude and zenith cut of the KASCADE array, the Galactic centre is never observed, and the exposure-weighted halo column is about two-thirds of the all-sky mean.

The absorption of Sec.~\ref{sec:gamma} is re-evaluated over the same exposure and applied inside the integral, energy by energy, and not as a single factor. This matters because the pair-production cross section on the microwave background peaks near $2\PeV$: the surviving fraction is smallest near the threshold and rises again above it, so a single factor taken at the threshold would understate the predicted flux at the energies where most of it lies.

Two ingredients of the KASCADE analysis we cannot reproduce. The two energies they tabulate are not thresholds but medians over the showers above the size cut, and their detection efficiency is averaged over simulations generated with an $E^{-2}$ spectrum. Using the median photon energy as the lower limit of our integral therefore treats a median as a threshold, and how their selection responds to a spectrum harder than the one it was tuned on is not published.  We can try to estimate the approximate size of both corrections. The limit scales inversely with the efficiency, and we vary the energy assignment by $30\%$ either way; together, they leave the exclusion between about $60$ and about $900$, thus the exclusion remains robust even after accounting for this uncertainty.
\bibliographystyle{JHEP}
\bibliography{refs}

@article{DAMPE:2019gys,
    author = "An, Q. and others",
    collaboration = "DAMPE",
    title = "{Measurement of the cosmic-ray proton spectrum from 40 GeV to 100 TeV with the DAMPE satellite}",
    eprint = "1909.12860",
    archivePrefix = "arXiv",
    primaryClass = "astro-ph.HE",
    doi = "10.1126/sciadv.aax3793",
    journal = "Sci. Adv.",
    volume = "5",
    number = "9",
    pages = "eaax3793",
    year = "2019"
}

@article{GRAPES-3:2024mhy,
    author = "Varsi, F. and others",
    collaboration = "GRAPES-3",
    title = "{Evidence of a Hardening in the Cosmic Ray Proton Spectrum at around 166~TeV Observed by the GRAPES-3 Experiment}",
    doi = "10.1103/PhysRevLett.132.051002",
    journal = "Phys. Rev. Lett.",
    volume = "132",
    number = "5",
    pages = "051002",
    year = "2024"
}

@article{Dondarini:2025ktz,
    author = "Dondarini, Alessandro and Marino, Giulio and Panci, Paolo and Zantedeschi, Michael",
    title = "{The fast, the slow and the merging: probes of evaporating memory burdened PBHs}",
    eprint = "2506.13861",
    archivePrefix = "arXiv",
    primaryClass = "hep-ph",
    doi = "10.1088/1475-7516/2025/11/006",
    journal = "JCAP",
    volume = "11",
    pages = "006",
    year = "2025"
}

@article{LHAASO:2025byy,
    author = "Cao, Zhen and others",
    collaboration = "LHAASO",
    title = "{Precise measurements of the cosmic ray proton energy spectrum in the ``knee'' region}",
    eprint = "2505.14447",
    archivePrefix = "arXiv",
    primaryClass = "astro-ph.HE",
    doi = "10.1016/j.scib.2025.10.048",
    journal = "Sci. Bull.",
    volume = "70",
    pages = "4173--4180",
    year = "2025"
}

@article{LHAASO:2025mlf,
    author = "Cao, Zhen and others",
    collaboration = "LHAASO",
    title = "{Precise Measurement of the Cosmic Ray Helium Spectrum above 0.1 PeV}",
    eprint = "2511.05013",
    archivePrefix = "arXiv",
    primaryClass = "astro-ph.HE",
    doi = "10.1103/d838-49gt",
    journal = "Phys. Rev. Lett.",
    volume = "136",
    number = "12",
    pages = "121001",
    year = "2026"
}

@article{Lv:2024wrs,
    author = "Lv, Xing-Jian and others",
    title = "{Precise Measurement of the Cosmic-Ray Spectrum and $\langle \ln A \rangle$ by LHAASO: Connecting the Galactic to the Extragalactic Components}",
    eprint = "2403.11832",
    archivePrefix = "arXiv",
    primaryClass = "astro-ph.HE",
    doi = "10.3847/1538-4357/ada426",
    journal = "Astrophys. J.",
    volume = "979",
    number = "2",
    pages = "225",
    year = "2025"
}

@article{LHAASO:2023gne,
    author = "Cao, Zhen and others",
    collaboration = "LHAASO",
    title = "{Measurement of Ultra-High-Energy Diffuse Gamma-Ray Emission of the Galactic Plane from 10 TeV to 1 PeV with LHAASO-KM2A}",
    eprint = "2305.05372",
    archivePrefix = "arXiv",
    primaryClass = "astro-ph.HE",
    doi = "10.1103/PhysRevLett.131.151001",
    journal = "Phys. Rev. Lett.",
    volume = "131",
    number = "15",
    pages = "151001",
    year = "2023"
}

@article{LHAASO:2022yxw,
    author = "Cao, Zhen and others",
    collaboration = "LHAASO",
    title = "{Constraints on Heavy Decaying Dark Matter from 570 Days of LHAASO Observations}",
    eprint = "2210.15989",
    archivePrefix = "arXiv",
    primaryClass = "astro-ph.HE",
    doi = "10.1103/PhysRevLett.129.261103",
    journal = "Phys. Rev. Lett.",
    volume = "129",
    number = "26",
    pages = "261103",
    year = "2022"
}

@article{Castro:2025wgf,
    author = "Castro, Luis Enrique Espinosa and others",
    title = "{LHAASO protons versus LHAASO diffuse gamma-rays: a consistency check}",
    eprint = "2506.06593",
    archivePrefix = "arXiv",
    primaryClass = "astro-ph.HE",
    doi = "10.1093/mnrasl/slaf085",
    journal = "Mon. Not. Roy. Astron. Soc. Lett.",
    volume = "543",
    pages = "L20--L26",
    year = "2025"
}

@article{Aharonian:2026tzf,
    author = "Aharonian, Felix and Zhang, Bing Theodore",
    title = "{A Minimal Interpretation of the Galactic Cosmic-Ray Proton and Helium Spectra from GeV to PeV Energies}",
    eprint = "2602.08223",
    archivePrefix = "arXiv",
    primaryClass = "astro-ph.HE",
    doi = "10.3847/1538-4357/ae81a7",
    journal = "Astrophys. J.",
    volume = "1006",
    number = "2",
    pages = "211",
    year = "2026"
}

@article{Chung:1998zb,
    author = "Chung, Daniel J. H. and Kolb, Edward W. and Riotto, Antonio",
    title = "{Superheavy dark matter}",
    eprint = "hep-ph/9802238",
    archivePrefix = "arXiv",
    primaryClass = "hep-ph",
    doi = "10.1103/PhysRevD.59.023501",
    journal = "Phys. Rev. D",
    volume = "59",
    pages = "023501",
    year = "1998"
}

@article{Chung:1998ua,
    author = "Chung, Daniel J. H. and Kolb, Edward W. and Riotto, Antonio",
    title = "{Nonthermal supermassive dark matter}",
    eprint = "hep-ph/9805473",
    archivePrefix = "arXiv",
    primaryClass = "hep-ph",
    doi = "10.1103/PhysRevLett.81.4048",
    journal = "Phys. Rev. Lett.",
    volume = "81",
    pages = "4048--4051",
    year = "1998"
}

@article{Kolb:1998ki,
    author = "Kolb, Edward W. and Chung, Daniel J. H. and Riotto, Antonio",
    title = "{WIMPzillas!}",
    eprint = "hep-ph/9810361",
    archivePrefix = "arXiv",
    primaryClass = "hep-ph",
    doi = "10.1063/1.59655",
    journal = "AIP Conf. Proc.",
    volume = "484",
    number = "1",
    pages = "91--105",
    year = "1999"
}

@article{Chung:2001cb,
    author = "Chung, Daniel J. H. and others",
    title = "{On the Gravitational Production of Superheavy Dark Matter}",
    eprint = "hep-ph/0104100",
    archivePrefix = "arXiv",
    primaryClass = "hep-ph",
    doi = "10.1103/PhysRevD.64.043503",
    journal = "Phys. Rev. D",
    volume = "64",
    pages = "043503",
    year = "2001"
}

@article{Kuzmin:1997jua,
    author = "Kuzmin, V. A. and Rubakov, V. A.",
    title = "{Ultrahigh-energy cosmic rays: A Window to postinflationary reheating epoch of the universe?}",
    eprint = "astro-ph/9709187",
    archivePrefix = "arXiv",
    primaryClass = "astro-ph",
    journal = "Phys. Atom. Nucl.",
    volume = "61",
    pages = "1028",
    year = "1998"
}

@article{Kuzmin:1998kk,
    author = "Kuzmin, Vadim and Tkachev, Igor",
    title = "{Matter creation via vacuum fluctuations in the early universe and observed ultrahigh-energy cosmic ray events}",
    eprint = "hep-ph/9809547",
    archivePrefix = "arXiv",
    primaryClass = "hep-ph",
    doi = "10.1103/PhysRevD.59.123006",
    journal = "Phys. Rev. D",
    volume = "59",
    pages = "123006",
    year = "1999"
}

@article{Berezinsky:1997hy,
    author = "Berezinsky, V. and Kachelriess, M. and Vilenkin, A.",
    title = "{Ultrahigh-energy cosmic rays without GZK cutoff}",
    eprint = "astro-ph/9708217",
    archivePrefix = "arXiv",
    primaryClass = "astro-ph",
    doi = "10.1103/PhysRevLett.79.4302",
    journal = "Phys. Rev. Lett.",
    volume = "79",
    pages = "4302--4305",
    year = "1997"
}

@article{Kolb:2023ydq,
    author = "Kolb, Edward W. and Long, Andrew J.",
    title = "{Cosmological gravitational particle production and its implications for cosmological relics}",
    eprint = "2312.09042",
    archivePrefix = "arXiv",
    primaryClass = "astro-ph.CO",
    doi = "10.1103/RevModPhys.96.045005",
    journal = "Rev. Mod. Phys.",
    volume = "96",
    number = "4",
    pages = "045005",
    year = "2024"
}

@article{Griest:1989wd,
    author = "Griest, Kim and Kamionkowski, Marc",
    title = "{Unitarity Limits on the Mass and Radius of Dark Matter Particles}",
    doi = "10.1103/PhysRevLett.64.615",
    journal = "Phys. Rev. Lett.",
    volume = "64",
    pages = "615",
    year = "1990"
}

@article{Berezinsky:1998ed,
    author = "Berezinsky, V. and Kachelriess, M.",
    title = "{Limiting SUSY QCD spectrum and its application for decays of superheavy particles}",
    eprint = "hep-ph/9803500",
    archivePrefix = "arXiv",
    primaryClass = "hep-ph",
    doi = "10.1016/S0370-2693(98)00728-X",
    journal = "Phys. Lett. B",
    volume = "434",
    pages = "61--66",
    year = "1998"
}

@article{Bauer:2020jay,
    author = "Bauer, Christian W. and Rodd, Nicholas L. and Webber, Bryan R.",
    title = "{Dark matter spectra from the electroweak to the Planck scale}",
    eprint = "2007.15001",
    archivePrefix = "arXiv",
    primaryClass = "hep-ph",
    doi = "10.1007/JHEP06(2021)121",
    journal = "JHEP",
    volume = "06",
    pages = "121",
    year = "2021"
}

@article{Kachelriess:2018rty,
    author = "Kachelriess, M. and Kalashev, O. E. and Kuznetsov, M. Yu.",
    title = "{Heavy decaying dark matter and IceCube high energy neutrinos}",
    eprint = "1805.04500",
    archivePrefix = "arXiv",
    primaryClass = "astro-ph.HE",
    doi = "10.1103/PhysRevD.98.083016",
    journal = "Phys. Rev. D",
    volume = "98",
    number = "8",
    pages = "083016",
    year = "2018"
}

@article{Chianese:2021jke,
    author = "Chianese, Marco and others",
    title = "{Constraints on heavy decaying dark matter with current gamma-ray measurements}",
    eprint = "2108.01678",
    archivePrefix = "arXiv",
    primaryClass = "hep-ph",
    doi = "10.1088/1475-7516/2021/11/035",
    journal = "JCAP",
    volume = "11",
    pages = "035",
    year = "2021"
}

@article{Ishiwata:2019aet,
    author = "Ishiwata, Koji and others",
    title = "{Probing heavy dark matter decays with multi-messenger astrophysical data}",
    eprint = "1907.11671",
    archivePrefix = "arXiv",
    primaryClass = "astro-ph.HE",
    doi = "10.1088/1475-7516/2020/01/003",
    journal = "JCAP",
    volume = "01",
    pages = "003",
    year = "2020"
}

@article{Cohen:2016uyg,
    author = "Cohen, Timothy and others",
    title = "{$\gamma$-ray Constraints on Decaying Dark Matter and Implications for IceCube}",
    eprint = "1612.05638",
    archivePrefix = "arXiv",
    primaryClass = "hep-ph",
    doi = "10.1103/PhysRevLett.119.021102",
    journal = "Phys. Rev. Lett.",
    volume = "119",
    number = "2",
    pages = "021102",
    year = "2017"
}

@article{Kalashev:2016cre,
    author = "Kalashev, O. K. and Kuznetsov, M. Yu.",
    title = "{Constraining heavy decaying dark matter with the high energy gamma-ray limits}",
    eprint = "1606.07354",
    archivePrefix = "arXiv",
    primaryClass = "astro-ph.HE",
    doi = "10.1103/PhysRevD.94.063535",
    journal = "Phys. Rev. D",
    volume = "94",
    number = "6",
    pages = "063535",
    year = "2016"
}

@article{Rocamora:2025ddt,
    author = "Rocamora, Manuel and De La Torre Luque, Pedro and S\'anchez-Conde, Miguel A.",
    title = "{Constraints on ultra-heavy DM from TeV-PeV gamma-ray diffuse measurements}",
    eprint = "2509.09609",
    archivePrefix = "arXiv",
    primaryClass = "astro-ph.HE",
    doi = "10.1016/j.dark.2026.102322",
    journal = "Phys. Dark Univ.",
    volume = "52",
    pages = "102322",
    year = "2026"
}

@article{Boehm:2025qro,
    author = "Boehm, Celine and Laha, Ranjan and Maity, Tarak Nath",
    title = "{LHAASO galactic plane $\gamma$-rays strongly constrain heavy dark matter}",
    eprint = "2509.07982",
    archivePrefix = "arXiv",
    primaryClass = "hep-ph",
    doi = "10.1088/1475-7516/2026/09/032",
    journal = "JCAP",
    volume = "09",
    pages = "032",
    year = "2026"
}

@article{Dubey:2025ouh,
    author = "Dubey, Abhishek and Saha, Akash Kumar",
    title = "{Breaking Dark: Hunting Heavy Decaying Dark Matter with Tibet AS$_\gamma$ and LHAASO-KM2A}",
    eprint = "2509.08039",
    archivePrefix = "arXiv",
    primaryClass = "hep-ph",
    doi = "10.1103/sf6v-s35h",
    journal = "Phys. Rev. D",
    volume = "114",
    number = "2",
    pages = "023023",
    year = "2026"
}

@article{Das:2023wtk,
    author = "Das, Saikat and Murase, Kohta and Fujii, Toshihiro",
    title = "{Revisiting ultrahigh-energy constraints on decaying superheavy dark matter}",
    eprint = "2302.02993",
    archivePrefix = "arXiv",
    primaryClass = "astro-ph.HE",
    doi = "10.1103/PhysRevD.107.103013",
    journal = "Phys. Rev. D",
    volume = "107",
    number = "10",
    pages = "103013",
    year = "2023"
}

@article{Alcantara:2019sco,
    author = "Alcantara, Esteban and Anchordoqui, Luis A. and Soriano, Jorge F.",
    title = "{Hunting for superheavy dark matter with the highest-energy cosmic rays}",
    eprint = "1903.05429",
    archivePrefix = "arXiv",
    primaryClass = "hep-ph",
    doi = "10.1103/PhysRevD.99.103016",
    journal = "Phys. Rev. D",
    volume = "99",
    number = "10",
    pages = "103016",
    year = "2019"
}

@article{PierreAuger:2025jwt,
    author = "Halim, A. Abdul and others",
    collaboration = "Pierre Auger",
    title = "{Search for a diffuse flux of photons with energies above tens of PeV at the Pierre Auger Observatory}",
    eprint = "2502.02381",
    archivePrefix = "arXiv",
    primaryClass = "astro-ph.HE",
    doi = "10.1088/1475-7516/2025/05/061",
    journal = "JCAP",
    volume = "05",
    pages = "061",
    year = "2025"
}

@article{Aloisio:2025nts,
    author = "Aloisio, Roberto and Ambrosone, Antonio and Evoli, Carmelo",
    title = "{Constraining superheavy dark matter with the KM3-230213A neutrino event}",
    eprint = "2508.08779",
    archivePrefix = "arXiv",
    primaryClass = "astro-ph.HE",
    doi = "10.1103/rc2p-53yg",
    journal = "Phys. Rev. D",
    volume = "113",
    number = "4",
    pages = "043024",
    year = "2026"
}

@article{KASCADEGrande:2017vwf,
    author = "Apel, W. D. and others",
    collaboration = "KASCADE Grande",
    title = "{KASCADE-Grande Limits on the Isotropic Diffuse Gamma-Ray Flux between 100 TeV and 1 EeV}",
    eprint = "1710.02889",
    archivePrefix = "arXiv",
    primaryClass = "astro-ph.HE",
    doi = "10.3847/1538-4357/aa8bb7",
    journal = "Astrophys. J.",
    volume = "848",
    number = "1",
    pages = "1",
    year = "2017"
}

@article{Vernetto:2016alq,
    author = "Vernetto, Silvia and Lipari, Paolo",
    title = "{Absorption of very high energy gamma rays in the Milky Way}",
    eprint = "1608.01587",
    archivePrefix = "arXiv",
    primaryClass = "astro-ph.HE",
    doi = "10.1103/PhysRevD.94.063009",
    journal = "Phys. Rev. D",
    volume = "94",
    number = "6",
    pages = "063009",
    year = "2016"
}

@article{Strong:2007nh,
    author = "Strong, Andrew W. and Moskalenko, Igor V. and Ptuskin, Vladimir S.",
    title = "{Cosmic-ray propagation and interactions in the Galaxy}",
    eprint = "astro-ph/0701517",
    archivePrefix = "arXiv",
    primaryClass = "astro-ph",
    doi = "10.1146/annurev.nucl.57.090506.123011",
    journal = "Ann. Rev. Nucl. Part. Sci.",
    volume = "57",
    pages = "285--327",
    year = "2007"
}

@article{Genolini:2019ewc,
    author = "G\'enolini, Y. and others",
    title = "{Cosmic-ray transport from AMS-02 boron to carbon ratio data: Benchmark models and interpretation}",
    eprint = "1904.08917",
    archivePrefix = "arXiv",
    primaryClass = "astro-ph.HE",
    doi = "10.1103/PhysRevD.99.123028",
    journal = "Phys. Rev. D",
    volume = "99",
    number = "12",
    pages = "123028",
    year = "2019"
}

@article{Loeb:2002ee,
    author = "Loeb, Abraham and Waxman, Eli",
    title = "{Galactic constraints on the sources of ultra high-energy cosmic rays}",
    eprint = "astro-ph/0205272",
    archivePrefix = "arXiv",
    primaryClass = "astro-ph",
    year = "2002"
}

@article{Evoli:2016xgn,
    author = "Evoli, Carmelo and others",
    title = "{Cosmic-ray propagation with DRAGON2: I. numerical solver and astrophysical ingredients}",
    eprint = "1607.07886",
    archivePrefix = "arXiv",
    primaryClass = "astro-ph.HE",
    doi = "10.1088/1475-7516/2017/02/015",
    journal = "JCAP",
    volume = "02",
    pages = "015",
    year = "2017"
}

@article{Masip:2008mk,
    author = "Masip, Manuel and Mastromatteo, Iacopo",
    title = "{Cosmic-ray knee and diffuse gamma, anti-e and anti-p fluxes from collisions of cosmic rays with dark matter}",
    eprint = "0810.4468",
    archivePrefix = "arXiv",
    primaryClass = "hep-ph",
    doi = "10.1088/1475-7516/2008/12/003",
    journal = "JCAP",
    volume = "12",
    pages = "003",
    year = "2008"
}

@article{Barcelo:2009uy,
    author = "Barcelo, Roberto and Masip, Manuel and Mastromatteo, Iacopo",
    title = "{Cosmic ray knee and new physics at the TeV scale}",
    eprint = "0903.5247",
    archivePrefix = "arXiv",
    primaryClass = "hep-ph",
    doi = "10.1088/1475-7516/2009/06/027",
    journal = "JCAP",
    volume = "06",
    pages = "027",
    year = "2009"
}

@article{LHAASO:2024knt,
    author = "Cao, Zhen and others",
    collaboration = "LHAASO",
    title = "{Measurements of All-Particle Energy Spectrum and Mean Logarithmic Mass of Cosmic Rays from 0.3 to 30 PeV with LHAASO-KM2A}",
    eprint = "2403.10010",
    archivePrefix = "arXiv",
    primaryClass = "astro-ph.HE",
    doi = "10.1103/PhysRevLett.132.131002",
    journal = "Phys. Rev. Lett.",
    volume = "132",
    number = "13",
    pages = "131002",
    year = "2024"
}

@article{Masip:2009bk,
    author = "Masip, Manuel and Mastromatteo, Iacopo",
    title = "{Cosmic-ray knee and flux of secondaries from interactions of cosmic rays with dark matter}",
    eprint = "0904.0921",
    archivePrefix = "arXiv",
    primaryClass = "hep-ph",
    year = "2009"
}

@article{Tomozawa:2010we,
    author = "Tomozawa, Yukio",
    title = "{Cosmic rays from AGN, the knee energy mass scale and dark matter particles}",
    eprint = "1002.1327",
    archivePrefix = "arXiv",
    primaryClass = "astro-ph.HE",
    year = "2010"
}

@article{Ciafaloni:2010ti,
    author = "Ciafaloni, Paolo and Comelli, Denis and Riotto, Antonio and Sala, Filippo and Strumia, Alessandro and Urbano, Alfredo",
    title = "{Weak Corrections are Relevant for Dark Matter Indirect Detection}",
    eprint = "1009.0224",
    archivePrefix = "arXiv",
    primaryClass = "hep-ph",
    doi = "10.1088/1475-7516/2011/03/019",
    journal = "JCAP",
    volume = "03",
    pages = "019",
    year = "2011"
}

@article{Weinrich:2020ftb,
    author = "Weinrich, N. and Boudaud, M. and Derome, L. and Genolini, Y. and Lavalle, J. and Maurin, D. and Salati, P. and Serpico, P. and Weymann-Despres, G.",
    title = "{Galactic halo size in the light of recent AMS-02 data}",
    eprint = "2004.00441",
    archivePrefix = "arXiv",
    primaryClass = "astro-ph.HE",
    doi = "10.1051/0004-6361/202038064",
    journal = "Astron. Astrophys.",
    volume = "639",
    pages = "A74",
    year = "2020"
}

@article{LHAASO:2025kyn,
    author = "Cao, Zhen and others",
    collaboration = "LHAASO",
    title = "{All-Sky Search for Individual Primordial Black Hole Bursts with LHAASO}",
    eprint = "2505.24586",
    archivePrefix = "arXiv",
    primaryClass = "astro-ph.HE",
    doi = "10.1103/nkby-9cs3",
    journal = "Phys. Rev. Lett.",
    volume = "135",
    pages = "181005",
    year = "2025"
}

@article{Yang:2024vij,
    author = "Yang, Chen and Wang, Sai and Zhao, Meng-Lin and Zhang, Xin",
    title = "{Search for the Hawking radiation of primordial black holes: prospective sensitivity of LHAASO}",
    eprint = "2408.10897",
    archivePrefix = "arXiv",
    primaryClass = "astro-ph.HE",
    doi = "10.1088/1475-7516/2024/10/083",
    journal = "JCAP",
    volume = "10",
    pages = "083",
    year = "2024"
}

@article{LHAASO:2023lkv,
    author = "Cao, Zhen and others",
    collaboration = "LHAASO",
    title = "{Very high-energy gamma-ray emission beyond 10 TeV from GRB 221009A}",
    eprint = "2310.08845",
    archivePrefix = "arXiv",
    primaryClass = "astro-ph.HE",
    doi = "10.1126/sciadv.adj2778",
    journal = "Sci. Adv.",
    volume = "9",
    pages = "adj2778",
    year = "2023"
}

@article{LHAASO:2023kyg,
    author = "Cao, Zhen and others",
    collaboration = "LHAASO",
    title = "{A tera-electron volt afterglow from a narrow jet in an extremely bright gamma-ray burst}",
    eprint = "2306.06372",
    archivePrefix = "arXiv",
    primaryClass = "astro-ph.HE",
    doi = "10.1126/science.adg9328",
    journal = "Science",
    volume = "380",
    pages = "1390--1396",
    year = "2023"
}

@article{Gao:2023und,
    author = "Gao, Lin-Qing and Bi, Xiao-Jun and Li, Jun and Yao, Run-Min and Yin, Peng-Fei",
    title = "{Constraints on axion-like particles from the observation of GRB 221009A by LHAASO}",
    eprint = "2310.11391",
    archivePrefix = "arXiv",
    primaryClass = "astro-ph.HE",
    doi = "10.1088/1475-7516/2024/01/026",
    journal = "JCAP",
    volume = "01",
    pages = "026",
    year = "2024"
}

@article{Li:2024ivs,
    author = "Li, Jun and Bi, Xiao-Jun and Gao, Lin-Qing and Huang, Xiaoyuan and Yao, Run-Min and Yin, Peng-Fei",
    title = "{Constraints on Axion-like Particles from the Observation of Galactic Sources by LHAASO}",
    eprint = "2401.01829",
    archivePrefix = "arXiv",
    primaryClass = "astro-ph.HE",
    doi = "10.1088/1674-1137/ad361e",
    journal = "Chin. Phys. C",
    volume = "48",
    pages = "065107",
    year = "2024"
}

@article{LHAASO:2024lub,
    author = "Cao, Zhen and others",
    collaboration = "LHAASO",
    title = "{Stringent Tests of Lorentz Invariance Violation from LHAASO Observations of GRB 221009A}",
    eprint = "2402.06009",
    archivePrefix = "arXiv",
    primaryClass = "astro-ph.HE",
    doi = "10.1103/PhysRevLett.133.071501",
    journal = "Phys. Rev. Lett.",
    volume = "133",
    pages = "071501",
    year = "2024"
}

@article{Cao:2021tat,
    author = "Cao, Zhen and others",
    title = "{Exploring Lorentz Invariance Violation from Ultrahigh-Energy $\gamma$ Rays Observed by LHAASO}",
    eprint = "2106.12350",
    archivePrefix = "arXiv",
    primaryClass = "astro-ph.HE",
    doi = "10.1103/PhysRevLett.128.051102",
    journal = "Phys. Rev. Lett.",
    volume = "128",
    pages = "051102",
    year = "2022"
}

@article{Satunin:2025hbk,
    author = "Satunin, P. S. and Troitsky, S. V.",
    title = "{Testing New-Physics Scenarios with the Combined LHAASO and Carpet-3 Fluence Spectrum of GRB 221009A: Axion-Like Particles and Lorentz-Invariance Violation}",
    eprint = "2510.07234",
    archivePrefix = "arXiv",
    primaryClass = "astro-ph.HE",
    doi = "10.1134/S0021364025609297",
    journal = "JETP Lett.",
    volume = "123",
    pages = "73--79",
    year = "2026"
}

@article{Barman:2025gjr,
    author = "Barman, Basabendu and Das, Arindam and Sarmah, Prantik and SivaKumar, Rakesh Kumar",
    title = "{Constraining gravitational dark matter with LHAASO and Fermi-LAT}",
    eprint = "2512.09997",
    archivePrefix = "arXiv",
    primaryClass = "hep-ph",
    doi = "10.1103/j6pw-vm1c",
    journal = "Phys. Rev. D",
    volume = "114",
    pages = "023013",
    year = "2026"
}

@article{Li:2021duv,
    author = "Li, Chengyi and Ma, Bo-Qiang",
    title = "{LHAASO discovery of highest-energy photons towards new physics}",
    eprint = "2109.07794",
    archivePrefix = "arXiv",
    primaryClass = "hep-ph",
    doi = "10.1016/j.scib.2021.07.030",
    journal = "Sci. Bull.",
    volume = "66",
    pages = "2254--2256",
    year = "2021"
}

@article{Kaci:2025gyb,
    author = "Kaci, Samy and Giacinti, Gwenael and Aharonian, Felix and Wang, Jie-Shuang",
    title = "{Microquasars as the major contributors to Galactic cosmic rays around the ``knee''}",
    eprint = "2510.01369",
    archivePrefix = "arXiv",
    primaryClass = "astro-ph.HE",
    year = "2025"
}

@article{Zhang:2025tew,
    author = "Zhang, B. Theodore and Kimura, Shigeo S. and Murase, Kohta",
    title = "{Microquasar jet-cocoon systems as PeVatrons}",
    eprint = "2506.20193",
    archivePrefix = "arXiv",
    primaryClass = "astro-ph.HE",
    doi = "10.1103/p6r1-qg5q",
    journal = "Phys. Rev. D",
    volume = "112",
    pages = "123015",
    year = "2025"
}

@article{Zhang:2026igt,
    author = "Zhang, B. Theodore and Yu, Shiqi",
    title = "{Microquasar Remnants as Pevatrons Illuminating the Galactic Cosmic Ray Knee}",
    eprint = "2602.08940",
    archivePrefix = "arXiv",
    primaryClass = "astro-ph.HE",
    year = "2026"
}

@article{Qiu:2026kdu,
    author = "Qiu, Zijian and Lin, Sujie and Yang, Lili",
    title = "{Collective Winds of Massive Star Clusters as the Dominant PeVatrons for Galactic Cosmic Rays}",
    eprint = "2605.31362",
    archivePrefix = "arXiv",
    primaryClass = "astro-ph.HE",
    year = "2026"
}

@article{Evoli:2026rpj,
    author = "Evoli, Carmelo",
    title = "{A Population View of the Cosmic-Ray Knee: The Role of Variance in Supernova Maximum Rigidities}",
    eprint = "2608.15892",
    archivePrefix = "arXiv",
    primaryClass = "astro-ph.HE",
    doi = "10.53941/pac.2026.100008",
    journal = "Phys. Cosmos",
    volume = "1",
    pages = "8",
    year = "2026"
}

@article{EspinosaCastro:2026xlr,
    author = "Espinosa Castro, Luis E. and Murase, Kohta and Rizza, Carlo and Villante, Francesco L. and Vecchiotti, Vittoria and Pagliaroli, Giulia",
    title = "{Multimessenger Concordance for the Cygnus Region as the Source of the Cosmic-Ray Knee}",
    eprint = "2603.21665",
    archivePrefix = "arXiv",
    primaryClass = "astro-ph.HE",
    year = "2026"
}

@article{Shi:2026unk,
    author = "Shi, Zhaodong and Wang, Guangwei and Yang, Rui-zhi and Aharonian, Felix",
    title = "{Microquasar Cygnus X-3 as the PeVatron Powering the Cygnus Bubble}",
    eprint = "2607.07100",
    archivePrefix = "arXiv",
    primaryClass = "astro-ph.HE",
    doi = "10.3847/2041-8213/ae878e",
    journal = "Astrophys. J. Lett.",
    volume = "1006",
    pages = "L31",
    year = "2026"
}

@article{EspinosaCastro:2026xbs,
    author = "Espinosa Castro, Luis Enrique and Evoli, Carmelo and Blasi, Pasquale",
    title = "{Transition from Diffusion to Drift-Dominated Cosmic Ray Transport and the Origin of the Knee}",
    eprint = "2606.28141",
    archivePrefix = "arXiv",
    primaryClass = "astro-ph.HE",
    year = "2026"
}

@article{Yuan:2025xqt,
    author = "Yuan, Qiang",
    title = "{Implication of multiple source populations of Galactic cosmic rays from proton and helium spectra}",
    eprint = "2511.06733",
    archivePrefix = "arXiv",
    primaryClass = "astro-ph.HE",
    doi = "10.1103/pzxy-v9v8",
    journal = "Phys. Rev. D",
    volume = "113",
    pages = "123037",
    year = "2026"
}

@article{Dzhatdoev:2026fvw,
    author = "Dzhatdoev, Timur A. and Semenov, Anatoly A.",
    title = "{Elemental cosmic ray spectra reveal two populations of Galactic sources and an immediate transition to an extragalactic component after the knee}",
    eprint = "2606.02748",
    archivePrefix = "arXiv",
    primaryClass = "astro-ph.HE",
    year = "2026"
}

@article{Navarro:1996gj,
    author = "Navarro, Julio F. and Frenk, Carlos S. and White, Simon D. M.",
    title = "{A Universal density profile from hierarchical clustering}",
    eprint = "astro-ph/9611107",
    archivePrefix = "arXiv",
    primaryClass = "astro-ph",
    doi = "10.1086/304888",
    journal = "Astrophys. J.",
    volume = "490",
    pages = "493--508",
    year = "1997"
}

@article{Dvali:2020wft,
    author = "Dvali, Gia and Eisemann, Lukas and Michel, Marco and Zell, Sebastian",
    title = "{Black hole metamorphosis and stabilization by memory burden}",
    eprint = "2006.00011",
    archivePrefix = "arXiv",
    primaryClass = "hep-th",
    doi = "10.1103/PhysRevD.102.103523",
    journal = "Phys. Rev. D",
    volume = "102",
    number = "10",
    pages = "103523",
    year = "2020"
}

@article{Alexandre:2024nuo,
    author = "Alexandre, Ana and Dvali, Gia and Koutsangelas, Emmanouil",
    title = "{New mass window for primordial black holes as dark matter from the memory burden effect}",
    eprint = "2402.14069",
    archivePrefix = "arXiv",
    primaryClass = "hep-ph",
    doi = "10.1103/PhysRevD.110.036004",
    journal = "Phys. Rev. D",
    volume = "110",
    number = "3",
    pages = "036004",
    year = "2024"
}

@article{Chianese:2025wrk,
    author = "Chianese, Marco",
    title = "{High-energy gamma-ray emission from memory-burdened primordial black holes}",
    eprint = "2504.03838",
    archivePrefix = "arXiv",
    primaryClass = "astro-ph.HE",
    doi = "10.1103/cgbp-2hdy",
    journal = "Phys. Rev. D",
    volume = "112",
    number = "2",
    pages = "023043",
    year = "2025"
}

@article{Tan:2025vxp,
    author = "Tan, Xiu-hui and Zhou, Yu-feng",
    title = "{Probing memory-burdened primordial black holes with galactic sources observed by LHAASO}",
    eprint = "2505.19857",
    archivePrefix = "arXiv",
    primaryClass = "astro-ph.CO",
    doi = "10.1016/j.physletb.2026.140404",
    journal = "Phys. Lett. B",
    volume = "876",
    pages = "140404",
    year = "2026"
}

@article{DeLaTorreLuque:2022PeV,
    author = "De La Torre Luque, Pedro and Gaggero, Daniele and Grasso, Dario and Fornieri, Ottavio and Egberts, Kathrin and Steppa, Constantin and Evoli, Carmelo",
    title = "{Galactic diffuse gamma rays meet the PeV frontier}",
    eprint = "2203.15759",
    archivePrefix = "arXiv",
    primaryClass = "astro-ph.HE",
    doi = "10.1051/0004-6361/202243714",
    journal = "Astron. Astrophys.",
    volume = "672",
    pages = "A58",
    year = "2023"
}

@article{DeLaTorreLuque:2025CRSea,
    author = "De La Torre Luque, Pedro and Gaggero, Daniele and Grasso, Dario and Marinelli, Antonio and Rocamora, Manuel",
    title = "{The Cosmic-Ray sea explains the diffuse Galactic gamma-ray and neutrino emissions from GeV to PeV}",
    eprint = "2502.18268",
    archivePrefix = "arXiv",
    primaryClass = "astro-ph.HE",
    year = "2025"
}

@article{Kantzas:2023oww,
    author = "Kantzas, Dimitrios and Markoff, Sera and Cooper, Alex J. and Gaggero, Daniele and Petropoulou, Maria and De La Torre Luque, Pedro",
    title = "{Possible contribution of X-ray binary jets to the Galactic cosmic ray and neutrino flux}",
    eprint = "2306.12715",
    archivePrefix = "arXiv",
    primaryClass = "astro-ph.HE",
    reportNumber = "LAPTH-007/23",
    doi = "10.1093/mnras/stad1909",
    journal = "Mon. Not. Roy. Astron. Soc.",
    volume = "524",
    number = "1",
    pages = "1326--1342",
    year = "2023"
}

@article{Cooper:2020tzq,
    author = "Cooper, A. J. and Gaggero, D. and Markoff, S. and Zhang, S.",
    title = "{High-energy Cosmic Ray production in X-ray Binary Jets}",
    eprint = "2002.01477",
    archivePrefix = "arXiv",
    primaryClass = "astro-ph.HE",
    doi = "10.1093/mnras/staa373",
    journal = "Mon. Not. Roy. Astron. Soc.",
    volume = "493",
    number = "3",
    pages = "3212--3222",
    year = "2020"
}

\end{document}